\documentclass[prx,twocolumn,showpacs,amsmath,amssymb,superscriptaddress,floatfix,longbibliography,nofootinbib]{revtex4-2}
\usepackage{setspace}
\usepackage{grffile}
\usepackage{wrapfig}
\usepackage{rotating}
\usepackage[normalem]{ulem}
\usepackage{textcomp}
\usepackage{verbatim}
\usepackage{amsmath,amssymb,amsthm,amsfonts}
\usepackage{mathrsfs}
\usepackage{mathtools}
\usepackage{bm}
\usepackage{dsfont}
\usepackage{thmtools}
\usepackage{thm-restate}
\usepackage{graphicx}
\usepackage[colorlinks=true, allcolors=blue]{hyperref}
\usepackage{tikz}
\usetikzlibrary{quantikz}
\hypersetup{
    colorlinks=true, 
    linkcolor=cyan,
    citecolor=magenta, 
    filecolor=magenta, 
    urlcolor=cyan,
    runcolor=cyan
}
\usepackage[capitalise]{cleveref}

\def\>{\rangle}
\def\<{\langle}
\def\Tr{\mathrm{Tr}}

\def\poly{\mathrm{poly}}
\def\Im{\mathrm{Im}}
\newcommand{\ident}{\mathbb{I}}
\newcommand{\ketbra}[1]{|{#1}\>\!\<#1|}

\newcommand{\bk}[2]{\<{#1}|{#2}\>}
\newcommand{\brak}[2]{\<{#1}|{#2}\>}
\newcommand{\ketb}[2]{|{#1}\>\!\<#2|}

\newcommand{\expval}[2]{\<{#1}|{#2}|{#1}\>}

\newcommand{\mc}[1]{\mathcal{#1}}

\newcommand{\beq}{\begin{equation}}
\newcommand{\eeq}{\end{equation}}
\newcommand{\bes} {\begin{subequations}}
\newcommand{\ees} {\end{subequations}}

\newtheorem{mytheorem}{Theorem}
\newtheorem{mylemma}{Lemma}
\newtheorem{mycorollary}{Corollary}
\newtheorem{myproposition}{Proposition}

\newtheorem{mydefinition}{Definition}

\begin{document}
\title{Computational Complexity and Simulability of Non-Hermitian Quantum Dynamics} 
\author{Brian Barch}
\email[Current affiliation: \emph{Quantum and Condensed Matter Physics Group (T-4),
Los Alamos National Laboratory, Los Alamos, New Mexico 87545, USA}; email: ]{barch@lanl.gov}
\affiliation{Department of Physics and Astronomy, University of Southern California, Los Angeles, California 90089, USA}
\affiliation{Center for Quantum Information Science \& Technology, University of Southern California, Los Angeles, California 90089, USA}
\author{Daniel Lidar}
\affiliation{Department of Physics and Astronomy, University of Southern California, Los Angeles, California 90089, USA}
\affiliation{Center for Quantum Information Science \& Technology, University of Southern California, Los Angeles, California 90089, USA}
\affiliation{Department of Electrical and Computer Engineering, University of Southern California, Los Angeles, California 90089, USA}
\affiliation{Department of Chemistry, University of Southern California, Los Angeles, California 90089, USA}
\date{\today}

\begin{abstract}
Non-Hermitian (NH) quantum systems demonstrate striking differences from their Hermitian counterparts, leading to claims of NH advantage in areas ranging from metrology to entanglement generation. 
We show that in the context of quantum computation, any such NH advantage is unlikely to be scalable as an efficient computational resource: if coherent non-unitary evolution with renormalization could be realized with only polynomial overhead, then the resulting model could implement postselection, implying implausibly strong complexity-theoretic power under standard assumptions.
We define $\mathrm{NHBQP}(U)$ as the computational power of polynomial-size quantum circuits that,
in addition to a standard universal unitary gate set, may apply a fixed gate $U$ on $O(1)$ qubits that is not proportional to a unitary,
with the state renormalized after each use of $U$. 
We prove that this model is powerful enough to decide every language in 
$\mathrm{PostBQP}$ (equivalently $\mathrm{PP}$).
Moreover, in the standard uniform circuit-family model this characterization is tight: for any fixed such non-unitary gate $U$, $\mathrm{NHBQP}(U)=\mathrm{PostBQP}=\mathrm{PP}$. $\mathrm{PostBQP}$ is believed intractable, so this suggests that any scalable NH computational advantage must come with a compensating cost limiting its efficiency.
Additionally, we study simulation complexity of restricted classes of non-unitary systems by purifying them to postselected unitary evolution in a form preserving system locality. Using this framework, we show that unitary gates with postselection can simulate not only evolution under NH Hamiltonians but arbitrary quantum trajectories.
Any NH model whose purification lies in a strongly simulable unitary family (e.g., Clifford, matchgate, or low-bond-dimension tensor-network circuits) remains efficiently classically simulable, provided the relevant postselected events occur with probability $\Omega(2^{-\poly(n)})$.
Thus adding non-Hermiticity to a universal unitary system makes it infeasibly computationally powerful, while adding it to a strongly simulable system adds no computational power in this setting.
\end{abstract}

\maketitle

\section{Introduction}
\label{sec:Introduction}

When a quantum system undergoes unitary dynamics interspersed with measurements, retaining the full measurement record leads to \emph{conditional} evolution that is not captured by closed-system unitary dynamics, nor by unconditional open-system descriptions, which average over outcomes.
In the limit of continuous weak measurement with minimally perturbing results, the combined effect of measurement back-action and Hamiltonian evolution can be expressed as coherent evolution under an effective non-Hermitian (NH) Hamiltonian~\cite{ashida2020nonhermitian, ashida2018full}.
NH Hamiltonians also arise in other contexts, including proposals to treat NH dynamics as fundamental via generalized inner products~\cite{Bender_1998_real_spectra, Bender_2002_complex_extension, mostafazadeh_pseudo-hermiticity_2002, mostafazadeh_2003}.
Experimentally, NH behavior has been realized in a variety of platforms~\cite{wu2019observation, li2019observation, naghiloo2019EP}, with particularly direct implementations in optical settings with engineered loss~\cite{ruter2010PT, regensburger2012PT, zeuner2015topological, hodaei2015tunable}.

Work on NH dynamics has revealed a range of novel phenomena, including measurement-induced phase transitions~\cite{gopalakrishnan_entanglement_2021, matsumoto2020continuous, PhysRevB.98.205136, PhysRevX.9.031009, PhysRevB.99.224307, PhysRevB.100.134306, PhysRevX.10.041020, PhysRevLett.125.030505, PhysRevX.11.011030, agarwal2026recognizing, barch2023scrambling}, violations or modifications of standard locality bounds~\cite{barch2024locality, ashida2018full, matsumoto2020continuous}, and changes in computational power relative to standard unitary models~\cite{zhang2025observation, abrams1998nonlinear, Mochizuki2023, Aaronson:2005aa}.
Motivated in part by these effects, a number of potential advantages over strictly Hermitian dynamics have been proposed, in the form of quantum speed limit violation~\cite{bender2007faster, lu2022realizing}, faster generation \cite{li2023speeding} and longer preservation~\cite{chakraborty2019delayed, fring2019eternal} of entanglement, and enhanced quantum sensing near exceptional points~\cite{hodaei2017enhanced, chen2017EP, parto2025enhanced, wu2025enhanced}.
A central practical question is whether such NH advantages can persist in scalable implementations, i.e., whether the required NH resources can be efficiently realized to achieve a task with bounded error.
In this work we address this question in the computational setting using tools from complexity theory.

The class of decision problems efficiently solvable with bounded error by a universal quantum computer is called $\mathrm{BQP}$~\cite{bernsteinQuantumComplexityTheory1997}.
It has been shown that 
coherent non-unitary evolution under certain NH Hamiltonians, 
interpreted as state evolution via renormalization, can decide problems outside $\mathrm{BQP}$~\cite{zhang2025observation, Aaronson:2005aa, zhang2026power, zhang2025physics}.
Moreover, conditional NH evolution is commonly expressed via a normalized (nonlinear) equation~\cite{ashida2020nonhermitian}, which is operationally equivalent to renormalizing output probabilities during measurement.
This is suggestive, as it has been shown that broad classes of nonlinear modifications of quantum mechanics would enable efficient solutions to problems believed to be intractable, including NP-complete and $\#\mathrm{P}$-hard problems~\cite{abrams1998nonlinear}.

Unlike earlier results that relate non-unitarity and postselection primarily at the level of computational power, we give an explicit construction that implements a postselection gadget from any fixed-support non-unitary gate.\footnote{Specifically, not \emph{proportional} to a unitary.} 
Under additional implementability and gap assumptions, this extends to size-dependent families approaching unitarity.
We then use $\mathrm{NHBQP}(U)$ to denote the class of problems solvable by polynomial-size quantum circuits built from
a universal unitary gate set together with a fixed non-unitary gate $U$  on $O(1)$ qubits, where each non-unitary step is
followed by explicit renormalization, equivalent to renormalizing during measurement.
We show that this model can decide every language in $\mathrm{PostBQP}=\mathrm{PP}$ (defined in the next section).
This bound is tight: for any fixed non-unitary gate $U$, $\mathrm{NHBQP}(U)=\mathrm{PostBQP}=\mathrm{PP}$ (\cref{thm:NHBQP-characterization}).\footnote{If one additionally allows a classical polynomial-time controller with adaptive oracle access to such postselected subroutines, the corresponding oracle class is $\mathrm{P}^{\mathrm{PP}}=\mathrm{P}^{\#\mathrm{P}}$, which contains the polynomial hierarchy by Toda's theorem~\cite{toda1991}. We do not assume such oracle access in the definition of $\mathrm{NHBQP}(U)$.}

As this level of computational power is widely regarded as implausible for physically realizable devices, we interpret our results as evidence that any scalable NH resource with the potential to enhance universal computation must be accompanied by a compensating cost---most naturally an exponentially small success probability---so that the overall implementation is not efficient.
This motivates careful resource accounting when interpreting NH-enabled protocols as scalable computational primitives, that is, claims of computational advantage should explicitly include the physical overhead of realizing the conditioning/renormalization, rather than only the normalized (postselected) dynamics.

In the second half of the paper, we take a complementary approach and study when NH dynamics remains efficiently classically simulable.
We do this by purifying non-unitary evolution into unitary evolution on a larger system followed by postselection.
This yields a simple criterion: if the purified unitary circuit family is strongly simulable, then the induced non-unitary conditional dynamics is also strongly simulable, provided the postselected events occur with probability $\Omega(2^{-\poly(n)})$.
Using this framework, we bound the simulation complexity of non-unitary extensions of paradigmatic strongly simulable families (Clifford and matchgate circuits), and we extend the purification approach to Trotterized simulations of general quantum trajectories in a locality-preserving form relevant for tensor-network methods.
These results help bound when non-Hermiticity is expected to increase computational power versus when it does not.

We begin with background on NH quantum mechanics, computational complexity, and simulability in \cref{sec:BG}.
In \cref{sec:NHBQP} we show how non-unitary gates implement postselection when supplemented by suitable unitaries, and we state the corresponding hardness consequences.
Extensions of simulability to NH systems and non-unitary circuits are discussed in \cref{sec:upperbounds}.
We conclude in \cref{sec:conclusion} with implications and directions for future work.
Technical calculations supporting several results are collected in the Appendix.

\section{Background}
\label{sec:BG}

\subsection{Non-Hermitian Quantum Mechanics}
\label{sec:BG-NH}

The paradigmatic conditional evolution motivating NH quantum mechanics is the \emph{no-jump} trajectory. This describes evolution in a setting where quantum jumps are possible, but where one conditions on the event that no jump occurs, either by chance or by postselection~\cite{ashida2020nonhermitian,brunsimplemodelquantum2002}. Consider a state $\rho$ evolving under the Lindblad equation~\cite{lindbladGeneratorsQuantumDynamical1976,alickiQuantumDynamicalSemigroups2007}
\beq
\label{eq:lindblad}
\dot \rho = -i[H_0,\rho]+\sum_a \left( L_a \rho L_a^\dag - \frac{1}{2}\{L_a^\dag L_a, \rho\} \right).
\eeq
In the quantum-trajectories picture, the jump term $L_a \rho L_a^\dag$ transfers weight between different trajectories, while the remaining terms generate the drift evolution within a single trajectory~\cite{breuerTheoryOpenQuantum2002}. Conditioning on the event that no jump occurs removes the jump term, yielding an unnormalized conditional evolution of the form
\beq
\dot \rho = -i\left(H\rho-\rho H^\dagger\right),
\eeq
with the effective NH Hamiltonian
\beq
\label{eq:H_eff}
H = H_0 -\frac{i}{2} \sum_a L_a^\dag L_a.
\eeq
See, e.g., \cref{eq:drift-evolution} and \cref{app:trajectories-purified} for a concrete example. The (unnormalized) no-jump map over time $t$ is
$\rho \mapsto U \rho U^\dag$ with $U=e^{-iHt}$.
This map is completely positive (CP) but not trace preserving (TP), and it is coherent, in that it maps pure states to pure states up to normalization. In the trajectories interpretation, $\Tr[U\rho U^\dag]$ is the probability of the no-jump measurement record, so it must lie in $[0,1]$.

Writing $H=H_0-i\Gamma$ with $\Gamma=\frac{1}{2}\sum_a L_a^\dag L_a\ge 0$, the trace of the unnormalized conditional state decreases according to
\beq
\label{eq:trace_decay}
\partial_t \Tr[\rho_t] = -2 \Tr[\Gamma \rho_t].
\eeq
Thus $\Gamma\ge 0$ is necessary and sufficient to ensure that, for every initial state, the no-jump probability $\Tr[\rho_t]$ is nonincreasing in time and hence never exceeds one.
Later we will frequently work with \emph{normalized} conditional states, in which overall rescalings of $U$ (equivalently, imaginary shifts of $H$ by multiples of $i\ident$) do not affect the dynamics.

There are two common approaches to studying NH dynamics: (i) normalized evolution of the conditional state, where $H$ represents effective conditional dynamics and is directly physically applicable, or (ii) the metric formalism underlying $PT$-symmetric (more generally quasi-Hermitian) systems, where $H$ is treated as fundamental in a modified Hilbert space~\cite{ashida2020nonhermitian,mostafazadeh_pseudo-hermiticity_2002,mostafazadeh_2003}. We primarily use the former approach, but will use techniques from the latter when it is convenient.

\subsubsection{Normalized evolution}
\label{sec:norm-evol}

To interpret non-unitary evolution as mapping states to states, one can normalize its action on pure states,
\beq
\label{eq:pure-evolution}
\ket{\psi_t} \equiv \frac{U\ket{\psi_0}}{\left\| U\ket{\psi_0}\right\|},
\eeq
or trace-normalize its action on mixed states,
\beq
\label{eq:mixed-evolution}
\rho_t \equiv \frac{U \rho_0 U^\dag}{\Tr[U \rho_0 U^\dag]}.
\eeq
This is the usual conditional-state update: the denominator is the Born-rule probability of the no-jump measurement record, and can be viewed as the normalization factor in Bayes rule~\cite{ashida2018full}. The same normalized evolution is obtained whether one normalizes the state continuously in time, normalizes the state after each discrete non-unitary step, or normalizes the output probabilities during measurement. 
In the continuous picture, differentiating \cref{eq:mixed-evolution} yields a nonlinear (quadratic) equation of motion
\beq
\label{eq:nonlinear_norm_eom}
\dot\rho = -i\left(H\rho-\rho H^\dag\right) - i\rho\Tr \left[\rho\left(H^\dag-H\right)\right].
\eeq
Introducing a constant imaginary shift $H\mapsto H+ic\ident$ rescales $U(t)=e^{-iHt}$ by $e^{ct}$ and therefore has no effect on the normalized maps \cref{eq:pure-evolution,eq:mixed-evolution}.
Such a shift does, however, rescale the unnormalized no-jump probability $\Tr[U(t)\rho U(t)^\dag]$ by a factor $e^{2ct}$, and hence changes success probabilities (and physical overhead) whenever $U\rho U^\dag$ is interpreted as a postselected/no-jump branch of an underlying trace-preserving evolution.
Accordingly, when we focus solely on normalized trajectories or circuit models with explicit renormalization, we may choose $c$ for convenience and need not impose $\Gamma\ge 0$ as a separate restriction. Whenever success probabilities are operationally relevant, however, the choice of $c$ (and hence the absolute scale of $\Gamma$) must be tracked explicitly.
Throughout this paper, when working in the normalized-evolution picture we will use ``non-unitary'' to mean ``not proportional to a unitary,'' since proportional operators are equivalent under normalization.

\subsubsection{$PT$-symmetric systems}
\label{sec:PT-sym}

A Hamiltonian is said to be pseudo-Hermitian if there exists a Hermitian, invertible,\footnote{Generalizations which allow singular $\eta$ can arise for nondiagonalizable Hamiltonians at exceptional points, but for our purposes it is sufficient to restrict to invertible $\eta$ and (when needed) diagonalizable $H$.} generally non-unique, ``metric'' operator $\eta$ such that $H^\dag \eta = \eta H$~\cite{Bender_1998_real_spectra, Bender_2002_complex_extension, mostafazadeh_pseudo-hermiticity_2002}. In this case, eigenvalues of $H$ are either real or come in complex conjugate pairs. If, moreover, there exists a positive definite $\eta$, then $H$ is said to be quasi-Hermitian; in the $PT$-symmetric literature this corresponds to the unbroken $PT$-symmetry phase.\footnote{Strictly, unbroken $PT$ symmetry is a sufficient condition for quasi-Hermiticity but not an equivalent definition.} 
We use the $PT$-symmetric terminology (rather than quasi-Hermitian) because it is more common in the physics literature for describing such systems. In this case one can write
\beq
\label{eq:SHS}
H = S H_0 S^{-1}
\eeq
for $S$ and $H_0$ both Hermitian, where $H_0$ is isospectral to $H$ under the similarity transform $S = \eta^{-1/2}$ (the positive square root of $\eta^{-1}$)~\cite{matsumoto2020continuous, karuvade2022observing}.
The associated evolution operator decomposes as $U = S U_0 S^{-1}$ with unitary $U_0=e^{-iH_0t}$. 
While $U$ is non-unitary in the standard inner product, one can use $\eta>0$ to define a $\eta$-inner product in which $H$ is self-adjoint and $U$ is norm-preserving. We will use this metric picture for geometric intuition, but otherwise work in the standard inner product.
Additionally notice that any diagonalizable pseudo-Hermitian $H = H_0 -i\Gamma$ necessarily has indefinite $\Gamma$, and so can only be physically generated up to an overall imaginary shift. This is shown in \cref{app:PT-indefinite-gamma}.
Intuitively, pseudo-Hermiticity forces the spectrum to be closed under complex conjugation, whereas a semidefinite $\Gamma$ would force all eigenvalues to have imaginary parts of the same sign; the only consistent possibility for a diagonalizable pseudo-Hermitian Hamiltonian with semidefinite $\Gamma$ is therefore $\Gamma=0$.

\subsubsection{Diagonalizing $H$}

When $H$ is diagonalizable but NH, it admits a biorthogonal spectral decomposition in terms of right and left eigenvectors $\{\ket{r_i}\}$ and $\{\bra{l_i}\}$,
\beq
\label{eq:diagonal}
H = \sum_i \nu_i \ketb{r_i}{l_i},
\eeq
with (generally complex) eigenvalues $\{\nu_i\}$. The eigenvectors can be chosen biorthonormal so that $\bk{l_i}{r_j}=\delta_{ij}$. One may additionally normalize $\bk{r_i}{r_i}=1$, at the cost of leaving the left eigenvectors $\{\bra{l_i}\}$ unnormalized. With this choice, the evolution operator is
\beq
U = \sum_i e^{-it \nu_i} \ketb{r_i}{l_i}.
\eeq
Real eigenvalues generate oscillatory behavior, while complex eigenvalues cause exponential growth or decay along the corresponding right-eigenspaces. 

When the imaginary parts are not all equal, any initial state undergoing normalized evolution will exponentially decay into subspace of right-eigenspaces whose eigenvalues have maximal imaginary component and on which it has nonzero support~\cite{barch2023scrambling}. That is, if $\mathscr{H}_\rho$ is the subspace where $\rho$ has support, $\nu^I_\mathrm{max}$ is the maximal imaginary eigenvalue component within that subspace, and $\mathscr{H}_L$ the long-time subspace which $\rho$ approaches, then
\begin{align}
\begin{split}
    \mathscr{H}_\rho &= \mathrm{span}\left\{\ket{r_j}\ \vert\ \bra{l_j}\rho\ket{l_j}>0\right\}\\
    \mathscr{H}_L &= \mathrm{span}\left\{\ket{r_j}\in\mathscr{H}_\rho\ \vert\ \Im({\nu_j}) = \nu_\mathrm{max}^I\right\}
\end{split}
\end{align}
For nondegenerate $H$ this subspace will be one dimensional, so such Hamiltonians are said to be in the ``purifying'' phase~\cite{gopalakrishnan_entanglement_2021}.

\subsection{Computational Complexity}
\label{sec:complex}

Computational complexity theory provides a framework for classifying computational tasks according to the resources required to carry them out, such as time or memory~\cite{Papadimitriou:book,Kitaev:book,arora_computational_2009}. Here we primarily focus on decision problems, where (informally) an algorithm must output a single bit representing ``yes'' or ``no'' given an input string. More formally, the algorithm must determine whether a given string belongs to a specified set of strings, a task referred to as \emph{deciding a language}. We also reference counting problems, where an algorithm must output the number of solutions associated with a given input, rather than determining if one exists. A complexity class is then defined as a collection of such languages (informally, computational problems) that can be decided (solved) within a given resource bound by a specified computational model.

We call a family of algorithms $\{C_n\}$ efficient if $C_n$ achieves a task using resources that scale polynomially with input size $n$. Moreover, the family is said to be uniform if there exists a classical algorithm that outputs a description of $C_n$ in time $\poly(n)$.
The class Polynomial time ($\mathrm{P}$) consists of all languages efficiently decidable by a uniform family of deterministic classical algorithms. Bounded-error Probabilistic Polynomial time ($\mathrm{BPP}$) extends this to classical randomized algorithms that output the correct answer with probability at least $2/3$, where any constant greater than $1/2$ can be amplified to $2/3$ by repetition and majority vote. The quantum analogue, $\mathrm{BQP}$, denotes the class of languages efficiently decidable with similar bounded error by a uniform family of quantum circuits. It is known that $\mathrm{P} \subseteq \mathrm{BPP} \subseteq \mathrm{BQP}$.

Several complexity classes believed to be intractable in general, but useful as reference points, include $\mathrm{PP}$, $\mathrm{\#P}$, and $\mathrm{PostBQP}$. $\mathrm{PP}$ can be viewed as a relaxation of $\mathrm{BPP}$ in which the acceptance probability need only be greater than $1/2$, potentially by an arbitrarily small margin.
Informally, $\mathrm{\#P}$ is a class of counting problems: given an input instance and an efficiently checkable yes/no condition, how many bitstrings satisfy that condition?\footnote{More precisely, a function $f$ is in $\mathrm{\#P}$ if there exists a polynomial $q$ and a polynomial-time computable predicate $R(x,w)$ such that
$f(x) = \left| \left\{ w\in\{0,1\}^{q(|x|)} : R(x,w)=1 \right\} \right|$.}
The two are related by the standard oracle equivalence $\mathrm{P}^{\#\mathrm{P}}=\mathrm{P}^{\mathrm{PP}}$ (see, e.g., ~\cite{Papadimitriou:book,arora_computational_2009}).
As usual, the oracle notation $\mathrm{P}^\mathrm{C}$ denotes the class solved by algorithms for $\mathrm{P}$ that may call a solver for C as an elementary subroutine.
$\mathrm{PostBQP}$ describes $\mathrm{BQP}$ enhanced by postselection, defined below.
It was shown that $\mathrm{PostBQP}=\mathrm{PP}$~\cite{Aaronson:2005aa}, hence
$\mathrm{P}^\mathrm{PostBQP} = \mathrm{P}^\mathrm{PP} = \mathrm{P}^\mathrm{\#P}$. Toda's theorem further shows that the entire polynomial hierarchy ($\mathrm{PH}$) is contained in $\mathrm{P}^{\#\mathrm{P}}$~\cite{toda1991}.

\subsubsection{Postselection}

Postselection is the hypothetical ability to condition on a desired measurement outcome after a computation has taken place~\cite{Aaronson:2005aa, boson-sampling-orig, aaronson2014postbqp, aaronson2014forrelation, aaronson2016, chen2016note}. More formally, $\mathrm{PostBQP}$ consists of decision problems solvable by a uniform family of polynomial-size quantum circuits with two designated output bits: an output bit and a postselection bit. The circuits must satisfy the following conditions on each input of size $n$:
(i) the postselection bit equals $1$ with probability at least $\Omega(2^{-\poly(n)})$, and
(ii) conditioned on the postselection bit being $1$, the output bit is correct with probability at least $2/3$~\cite{Aaronson:2005aa, aaronson2014postbqp}.
We use standard asymptotic notation: $P(s)\in\Omega(2^{-\poly(n)})$ means $P(s)\ge 2^{-p(n)}$ for some polynomial $p(n)$.\footnote{Specifically, $f(n) \in O(g(n))$ [$\Omega(g(n))$] if there exist positive constants $c$ and $n_0$ such that for all $n \ge n_0$: $0 \le f(n) \le c \cdot g(n)$ [$0 \le c \cdot g(n) \le f(n)$]. Also, $f(n) \in \Theta(g(n))$ if and only if $f(n) \in O(g(n))$ and $f(n) \in \Omega(g(n))$.}

When a circuit has multiple postselection bits in parallel, these can be combined into a single postselection bit by taking the conjunction using reversible AND gates; the postselection success probability then decreases multiplicatively, but remains at least $\Omega(2^{-\poly(n)})$ if only polynomially many bits are used. 
Intermediate, even interleaved, postselection steps within a single quantum computation do not increase the power of $\mathrm{PostBQP}$: one can coherently record each postselection condition in ancillas and postselect on their conjunction at the end of the computation~\cite{Aaronson:2005aa}.
By contrast, the oracle class $\mathrm{P}^{\mathrm{PostBQP}}=\mathrm{P}^{\mathrm{PP}}$ refers to classical polynomial-time computation with adaptive oracle access to a $\mathrm{PostBQP}=\mathrm{PP}$ subroutine (i.e., where later queries may depend on earlier oracle answers); whether this collapses to $\mathrm{PostBQP}$ is not known (the adaptive case is subtle because postselection can amplify even exponentially small error probabilities), and we will not rely on any separation here.

In the quantum circuit model, postselection is equivalent to applying a projector onto the subspace where the postselection bit is $\ket{1}$, followed by renormalization, prior to reading out the output bit. For our purposes, it will be important that exact postselection can be relaxed to an approximate version. Specifically, it suffices to implement a postselection gadget that is exponentially close to the ideal projector on instances where the postselected event has probability at least $\Omega(2^{-\poly(n)})$. We will show that such postselection gadgets can be implemented using normalized evolution under NH Hamiltonians.

\subsubsection{Simulability}
\label{sec:simulability}

Not all quantum circuit families realize the full computational power of $\mathrm{BQP}$; many have output statistics that can be efficiently reproduced by classical algorithms, in which case the circuit family is considered simulable. Here, we are primarily concerned with circuit families which are strongly simulable~\cite{bremner2011classical, nes2010classical, johnson2013solving}. Strong simulability requires a classical algorithm that can estimate marginal output probabilities to exponentially small additive error. We use the following strong-simulation notion~\cite{bremner2011classical}, which is the one relevant for postselection:

\begin{mydefinition}[Strong simulability]
\label{def:strong-simulability}
Let $\{C_n\}_{n\ge 1}$ be a uniform family of quantum circuits with a measured output register $X$
of length $\poly(n)$.
We say $\{C_n\}$ is \emph{strongly simulable} if there exists a classical algorithm that, given
$n$, a choice of indices $\mc{I}$ specifying any subset of the output bits, an assignment $x_{\mc{I}}$ to those
bits, and an error parameter $\epsilon>0$, outputs a probability estimate $\widetilde{P}(X_{\mc{I}}=x_{\mc{I}})$ satisfying 
\beq
\bigl|\widetilde{P}(X_{\mc{I}}=x_{\mc{I}})-P(X_{\mc{I}}=x_{\mc{I}})\bigr|\le \epsilon
\eeq
in time $\poly\bigl(n,\log(1/\epsilon)\bigr)$.
\end{mydefinition}

The need for exponentially small additive error becomes essential once one conditions on events that may occur with probability as small as $2^{-\poly(n)}$, as in postselection. The following proposition formalizes that strong simulability is preserved under such postselection.

\begin{myproposition}[Strong simulability is preserved under postselection]
\label{prop:strong-closed-under-postselection}
Let $\{C_n\}_{n\ge 1}$ be a family of circuits with measured output registers $(X,S)$, inducing a distribution $P_n(x,s)$ over bitstrings $(x,s)$, where $S$ is a designated postselection register and $X$ is a computational output register.
Assume $\{C_n\}$ is strongly simulable in the sense of \cref{def:strong-simulability}.

Fix a postselection string $s^\star$ such that there exists a polynomial $q(n)$ with $P_n(s^\star)\ge 2^{-q(n)}$ for all sufficiently large $n$.
Then any family of circuits which outputs the distribution
\beq
P_n^{\mathrm{post}}(x)\equiv P_n(x\mid s^\star)=\frac{P_n(x,s^\star)}{P_n(s^\star)}.
\eeq
is also strongly simulable. That is, for every $\epsilon>0$ there is a classical algorithm that outputs $\widetilde{P}_n^{\mathrm{post}}(x)$ satisfying $\bigl|\widetilde{P}_n^{\mathrm{post}}(x)-P_n^{\mathrm{post}}(x)\bigr|\le \epsilon$ in time $\poly\bigl(n,\log(1/\epsilon)\bigr)$.
\end{myproposition}

A proof is provided in \cref{app:strong-closed-under-postselection}. Note that the requirement of \emph{strong} simulation is essential here:
conditioning replaces probabilities by ratios such as $P_n(x,s^\star)/P_n(s^\star)$, and the
denominator $P_n(s^\star)$ may be as small as $2^{-\poly(n)}$.
A polynomial additive approximation to $P_n(x,s^\star)$ and $P_n(s^\star)$ is generally
insufficient, as one needs additive accuracy on the scale of $P_n(s^\star)$ in order to guarantee
constant error in the conditional probability.

Tensor networks are natural candidates for extending simulability results to the NH setting, since their contraction rules do not require unitarity, and for many local quantum systems they yield efficient classical simulation. For example, dynamical tensor-network methods can simulate 1D adiabatic and short-time unitary evolution while keeping bond dimensions polynomial via Lieb-Robinson bounds~\cite{osborne2006efficient,hastings2009quantum}. In NH systems, however, the effective notion of locality can change~\cite{barch2024locality}, making it unclear which tensor-network techniques (and bond-dimension guarantees) generalize.

More importantly for the present discussion, \cref{prop:strong-closed-under-postselection} highlights that accurate simulation of a postselected (and as we'll later show, purified-NH) model requires exponentially small (or zero) additive error on the relevant marginals. Tensor networks can, in principle, provide such strong simulation in regimes with sufficiently low entanglement~\cite{johnson2013solving}. In contrast, most practical approximate tensor-network time-evolution algorithms yield only polynomially small additive errors for many outcome probabilities, which can be amplified under postselection and thus generally fall short of strong simulation. For this reason, tensor networks are primarily of theoretical interest for the strong-simulability questions we address, even though they remain extremely useful as practical tools.

\subsubsection{Complexity of NH systems}

Normalized evolution under certain NH Hamiltonians is known to solve decision problems beyond $\mathrm{BQP}$~\cite{zhang2025observation, Aaronson:2005aa}.
This is striking, as NH Hamiltonians are often used as effective descriptions of open quantum systems, yet unconditional (e.g. Markovian) open-system evolution 
is described by CPTP maps and can be simulated by standard quantum circuits (and hence within the usual sampling model $\mathrm{SampBQP}$, the class of distributions samplable by polynomial-size quantum circuits \emph{without} postselection)~\cite{aloisio2023sampling, aaronson2010equivalence}.

The distinction is the conditioning step. Operationally, a non-unitary update arises as a Kraus branch $\rho\mapsto K\rho K^\dagger$ of a CPTP dilation, with success probability $p=\Tr[K\rho K^\dagger]$ and postselected state $\rho' = K\rho K^\dagger/p$. Without conditioning (i.e., when the measurement outcome is not fixed), the overall evolution is CPTP and does not exceed the power of ordinary quantum computation.
However, when the conditioned branch is treated as the output state, producing its output statistics amounts to sampling from a \emph{postselected} circuit.
Such postselected access can implement $\mathrm{PostBQP}=\mathrm{PP}$ computations~\cite{Aaronson:2005aa}. Accordingly, a generic polynomial-overhead simulation of these conditioned output distributions by ordinary $\mathrm{SampBQP}$ circuits would imply a collapse of postselection power (e.g., $\mathrm{PostBQP}=\mathrm{BQP}$), which is believed to be unlikely. Without such a collapse, one expects that reproducing the conditioned branch requires explicit postselection or potentially exponential repetition-until-success overhead.

Equivalently, the normalized non-unitary map is trace-preserving but nonlinear in $\rho$~\cite{ashida2020nonhermitian, ashida2018full}.
It was shown that broad classes of nonlinear modifications of quantum dynamics 
enable polynomial-time algorithms for $\mathrm{NP}$-complete decision problems and $\#\mathrm{P}$-complete counting problems~\cite{abrams1998nonlinear}. This motivates investigating which normalized NH dynamics can effectively implement the kind of conditioning needed for postselection-type power.

In this work we are specifically interested in coherent conditional circuits, which can be written in terms of non-unitary gates. 
Throughout, we work in the standard uniform quantum circuit model with polynomial-size circuits, renormalization after each non-unitary operation (equivalently, 
after conditioning on measurement outcomes), and standard gate-description assumptions (e.g., gate matrix entries that are efficiently approximable to inverse-exponential precision).

When gate entries are additionally assumed to be algebraic numbers,\footnote{More generally, one may assume a standard ``tameness''/approximability condition ensuring that nonzero postselection probabilities are at least $2^{-\poly(n)}$; see, e.g., Ref.~\cite{kuperberg_how_2014} for a detailed discussion in the $\mathrm{PostBQP}$ setting.} 
the set of languages decidable by polynomial-size non-unitary circuits with explicit renormalization is a subset of 
$\mathrm{PP}$~\cite{Aaronson:2005aa}. This can be seen by extending the proof that $\mathrm{BQP}\subseteq\mathrm{PP}$
to the present renormalized non-unitary setting~\cite{Aaronson:2005aa,adleman1997quantum}. As we show in \cref{sec:upperbounds}, it can also be shown constructively:
after a harmless overall rescaling that leaves the normalized map unchanged, each non-unitary step can be dilated into a unitary circuit on a larger system followed by postselection on a meter outcome, putting the resulting computational power within $\mathrm{PostBQP}$. 
Additionally, one can also express the acceptance probability of such circuits in a form reducible to $\mathrm{P}^{\#\mathrm{P}}$ computation, yielding a (looser) counting-complexity upper bound~\cite{zhang2026power}.

\section{Hardness of NH Hamiltonians + $\mathrm{BQP}$}
\label{sec:NHBQP}

Henceforth we work in the normalized-evolution picture, following every non-unitary operation with renormalization. All results in this section are statements about this formal renormalized circuit model; they do not depend on a particular physical interpretation of the renormalization step (e.g., as a normalized map induced by postselection/no-jump conditioning, versus treating renormalization as an abstract normalization rule).
Our goal in this section is to show that any fixed-support non-unitary gate can be used to implement a postselection gadget when supplemented by a universal set of unitary gates.
This implies that NH dynamics supplemented with universal unitary gates has computational power far beyond what is believed feasible, and is therefore unlikely to be scalable without an accompanying unacceptable resource cost.

\subsubsection{Single-qubit example}

The basic NH $\implies$ postselection mechanism is already apparent for the one-qubit NH Hamiltonian $H=i\sigma^z$:
\begin{align}
\begin{split}
e^{\sigma^z t}\bigl(a\ket{0}+b\ket{1}\bigr)
&= a e^{t}\ket{0}+b e^{-t}\ket{1}\\
&\sim \ket{0}+\frac{b}{a}e^{-2t}\ket{1},
\end{split}
\end{align}
where $\sim$ denotes equality up to state renormalization. 
For $t$ growing polynomially with the input size $n$, the $\ket{1}$ component is suppressed as $e^{-2t}=2^{-\poly(n)}$, so 
for inputs with non-negligible overlap on $\ket{0}$ ($|a|^2\ge 2^{-\poly(n)}$) the normalized map approximates a projector onto $\ket{0}$. This implements a postselection-type primitive, and when combined with universal unitary gates to implement the surrounding computation, yields computational power beyond $\mathrm{BQP}$ (made precise below).
This already suggests that if such evolution could be efficiently implemented for times $t\in\poly(n)$, then it would enable postselection-type power and hence implausibly strong complexity-theoretic consequences.

\subsection{Postselection and Hardness for Constant Support-Size Circuits}

We now demonstrate that this postselection mechanism is completely general: any non-unitary gate yields a postselection gadget when surrounded by suitable unitaries. We begin by formally defining the complexity class associated with $\mathrm{BQP}$ enhanced with any non-unitary gate.

\begin{figure}
\centering
\begin{quantikz}
\lstick{$\ket{0^{m-1}}$}&\qwbundle{m-1}&\gate[wires=2]{D=V^{\dag}UW}&\qw\rstick{$\ket{0^{m-1}}$}\\
\lstick{}&\qw &\qw &\qw\\
\end{quantikz}
\caption{A postselection gadget, which uses an $m$-qubit non-unitary gate $U$ together with unitaries $V,W$ from its SVD to implement an effective diagonal non-unitary map $D=V^\dag U W$ on a target qubit plus $m-1$ ancillas initialized to $\ket{0^{m-1}}$. This implements postselection on the target (bottom) qubit. Since $D$ is diagonal in the computational basis, the ancillas remain in the state $\ket{0^{m-1}}$.}
\label{fig:m-qubits-1-postselection}
\end{figure}

\begin{mydefinition}[$\mathrm{NHBQP}$ for a fixed non-unitary gate]
\label{def:NHBQP}
Fix a universal set of unitary gates and a gate $U$ acting on $m=O(1)$ qubits that is not proportional to a unitary.
Define $\mathrm{NHBQP}(U)$ as the class of languages decided with bounded error by a uniform family of polynomial-size circuits over this gate set, where each use of $U$ is followed by renormalization.
\end{mydefinition}

We assume the fixed gate set has efficiently computable algebraic matrix entries, as is standard in quantum complexity theory.
This avoids pathological encodings of non-uniform advice into arbitrary real constants and ensures that the counting-complexity characterizations used in \cref{sec:upperbounds} applies cleanly. For physical gates whose exact matrix entries are not algebraic (e.g., $U=e^{-iHt}$ at generic real times), one can work within this framework by fixing an algebraic approximation of $U$ to the required inverse-exponential precision.

We restrict attention to \emph{well-formed} circuits, meaning that on every input the renormalization after each use of $U$ is defined (equivalently, the postselected branch has nonzero norm at every such step). Circuits that violate this condition have undefined normalized evolution on that input and are not regarded as deciding a language with bounded error in this model.

\begin{mytheorem}[Constant-support non-unitary gates enable postselection]
\label{thm:NH-universality}
Let $U$ be a fixed non-unitary gate as in \cref{def:NHBQP} which can be implemented efficiently, and suppose each use of $U$ is followed by renormalization.
Then, together with any universal set of unitary gates, $U$ allows efficient implementation of a
postselection gadget with approximation error at most $2^{-\poly(n)}$ [i.e., conditioned on
successful postselection the probability of the undesired postselected outcome is 
$O(2^{-\poly(n)})$].
Consequently,
\beq
\mathrm{PostBQP}=\mathrm{PP} \subseteq \mathrm{NHBQP}(U).
\eeq
\end{mytheorem}

\begin{proof}
Write the singular value decomposition
\beq
U=VDW^\dagger,
\eeq
where $V,W$ are unitary and $D\ge 0$ is diagonal in a bitstring basis $\{\ket{s_j}\}$:
\beq
D=\sum_{j=1}^{d}\lambda_j\ketbra{s_j},
\quad
\lambda_1\ge\cdots\ge\lambda_{d}\ge 0,
\quad
d\equiv 2^m.
\eeq
Since $m=O(1)$ is constant, $V^\dag$ and $W$ are fixed matrices independent of $n$. They can be precomputed once and approximated to accuracy $2^{-\poly(n)}$ using $\poly(n)$ unitary gates from any universal set~\cite{nielsen_quantum_2010}. Implementing $D=V^\dag UW$ therefore incurs only polynomial overhead, even when the resulting gadget is repeated $r=\poly(n)$ times.

Because $U$ is not proportional to a unitary, it has at least two distinct singular values, so $\lambda_1>\lambda_{d}$. 
By permuting the diagonal entries of $D$ (absorbing the permutation into $V^\dag,W$), we may assume the largest singular value corresponds to the desired state $\ket{0^m}$ and the smallest to the undesired state $\ket{0^{m-1}1}$. Applying $D$ to the target qubit together with $m-1$ ancillas initialized to $\ket{0^{m-1}}$ ensures the evolution begins and ends in the two-dimensional span of $\ket{0^m}$ and $\ket{0^{m-1}1}$; see \cref{fig:m-qubits-1-postselection}. 
In this block, for any amplitudes $a,b\in\mathbb{C}$ and target qubit initialized to $a\ket{0}+b\ket{1}$ we have
\beq
\label{eq:D^r}
D^r\bigl(a\ket{0^m}+b\ket{0^{m-1}1}\bigr)
=
a\lambda_1^r\ket{0^m}+b\lambda_{d}^r\ket{0^{m-1}1}.
\eeq

After renormalization, measuring the target qubit yields the undesired outcome $1$ with probability
\beq
\label{eq:err}
P(\ket{0^{m-1}1})
=
\frac{|b|^2\lambda_{d}^{2r}}{|a|^2\lambda_1^{2r}+|b|^2\lambda_{d}^{2r}}
\le
\frac{1}{|a|^2}\left(\frac{\lambda_{d}}{\lambda_1}\right)^{2r}.
\eeq
We take this failure probability as the postselection-gadget approximation error.

The definition of $\mathrm{PostBQP}$ assumes the outcome being postselected occurs with probability at least $2^{-\poly(n)}$. In the present two-level reduction this corresponds to assuming
\beq
\label{eq:a-bound}
|a|^2\ge 2^{-k(n)}
\eeq
for some polynomial $k(n)$. To ensure $P(\ket{0^{m-1}1})\le \varepsilon$ for a given $\varepsilon$ uniformly over all inputs satisfying \cref{eq:a-bound}, it suffices to require
\beq
\label{eq:bound1}
\left(\frac{\lambda_{d}}{\lambda_1}\right)^{2r}\le \varepsilon 2^{-k(n)}.
\eeq

Define the normalized singular radius
\beq
\Delta\equiv 1-\frac{\lambda_{d}}{\lambda_1}\in(0,1].
\eeq
Then \cref{eq:bound1} is equivalent to
\beq
2r\bigl(-\ln(1-\Delta)\bigr)\ge \ln(2^{k(n)}/\varepsilon).
\eeq
Using the bound $-\ln(1-\Delta)\ge \Delta$,\footnote{This bound holds for $\Delta\in(0,1)$, but if $\Delta=1$ then $\lambda_d=0$ and the gadget is exact for $r=1$.} this is satisfied by any $r$ such that
\beq
\label{eq:r_suff}
r \ge \frac{\ln(2^{k(n)}/\varepsilon)}{2\Delta}.
\eeq
When $\varepsilon=2^{-p(n)}$ for some polynomial $p(n)$, \cref{eq:r_suff} can be satisfied by $r\in \Theta((p(n)+k(n))/\Delta)\in\poly(n)$, where the second inclusion holds since $\Delta$ is a constant. Therefore the implementation of a postselection gadget with the required error $2^{-\poly(n)}$ is efficient.

Succeeding a polynomial-depth circuit of universal unitary gates with a single postselection gadget creates a circuit which can decide any language in $\mathrm{PostBQP}$. Using $\mathrm{PostBQP}=\mathrm{PP}$~\cite{Aaronson:2005aa}, we conclude $\mathrm{PP}\subseteq \mathrm{NHBQP}(U)$.
\end{proof}

\subsection{Postselection and Hardness for Variable Support-Size Circuits}

When the non-unitary gate varies with input size $n$, we need to redefine the $\mathrm{NHBQP}$ complexity class.

\begin{mydefinition}[$\mathrm{NHBQP}$ for a family of non-unitary gates]
\label{def:NHBQP-family}
Fix a universal set of unitary gates. Let $\{U_n\}_{n\ge 1}$ be a family of non-unitary gates, where $U_n$ acts on $m(n)$ qubits with $m(n)\in O(\poly(n))$. 
Define $\mathrm{NHBQP}(\{U_n\})$ as the class of languages decided with bounded error by a uniform family of polynomial-size circuits $\{C_n\}$ over this gate set, where $C_n$ may use the gate $U_n$ followed by renormalization.
\end{mydefinition}
Here, ``uniform'' means that there exists a classical polynomial-time algorithm that outputs a description of $C_n$. 
When we allow a size-dependent family of non-unitary gates $\{U_n\}_{n\ge 1}$, we assume it is specified uniformly and does not encode non-uniform advice. That is, that there exists a classical polynomial-time algorithm that, on input $1^n$, outputs a description of $U_n$, and likewise that any auxiliary descriptions used below (e.g., circuit decompositions or the SVD ingredients) must be generable uniformly in $n$. Without such a restriction one can encode arbitrary truth tables into $\{U_n\}$ (or into $\{V_n^\dag,W_n\}$), thereby trivializing the model~\cite{arora_computational_2009}.
Two independent issues now arise: (i) whether the SVD unitaries $V_n^\dag,W_n$ are efficiently implementable, and (ii) whether the normalized singular radius closes sufficiently slowly with $n$. The next corollary makes these requirements explicit and shows that, under them, the same postselection-gadget construction remains efficient.

\begin{mycorollary}[Families of non-unitary gates]
\label{corr:NH-universality-family-clean}
Let $\{U_n\}$ be a family of non-unitary gates, where $U_n$ acts on $m(n)$ qubits and each use is followed by renormalization.
Assume that for any $n$ and any desired inverse-exponential accuracy $2^{-p(n)}$ for polynomial $p$, one can efficiently implement unitary circuits that approximate $V_n^\dag$ and $W_n$ from an SVD $U_n=V_nD_nW_n^\dagger$ to error at most $2^{-p(n)}$ (e.g., in operator norm).
Let $\lambda_{1,n}\ge\cdots\ge\lambda_{d,n}$ be the singular values of $U_n$, and define
\beq
\Delta_n \equiv 1-\frac{\lambda_{d,n}}{\lambda_{1,n}}.
\eeq
If $\Delta_n\in\Omega(1/\poly(n))$, then $\{U_n\}$ together with universal unitary gates can efficiently implement a postselection gadget with error $2^{-\poly(n)}$, and hence decide every language in $\mathrm{PostBQP}$. Consequently,
\beq
\mathrm{PostBQP}=\mathrm{PP} \subseteq \mathrm{NHBQP}(\{U_n\}).
\eeq
\end{mycorollary}

\begin{proof}
The proof follows \cref{thm:NH-universality} with $\Delta$ replaced by $\Delta_n$ and with the compilation of $V^\dag,W$ replaced by the assumed efficient implementations of $V_n^\dag,W_n$.
For target error $\varepsilon=2^{-p(n)}$ and postselection-event probability at least $2^{-k(n)}$, the same calculation yields the sufficient condition
\beq
r \ge \frac{\ln(2^{k(n)}/\varepsilon)}{2\Delta_n}.
\eeq
If $\Delta_n\in\Omega(1/\poly(n))$, then $r\in\poly(n)$ is sufficient, and the postselection-gadget reduction remains efficient.
\end{proof}

The normalized singular radius $\Delta_n$ has a ``distance-from-unitarity'' interpretation. Up to constant factors, $\Delta_n$ equals the minimum operator-norm distance between a rescaling of $U_n$ and the unitary group. In \cref{app:unitary-distance} we show that
\beq
\min_{\alpha>0,T\in\mathcal{U}(d)}\Vert \alpha U_n-T\Vert
=
\frac{\lambda_{1,n}-\lambda_{d,n}}{\lambda_{1,n}+\lambda_{d,n}}
=
c_n\Delta_n,
\eeq
where $c_n\equiv \lambda_{1,n}/(\lambda_{1,n}+\lambda_{d,n})\in(1/2,1]$.
Thus $\Delta_n\in\Omega(1/\poly(n))$ is equivalent to requiring that no rescaling $\alpha U_n$ is super-polynomially close to the unitary group in operator norm. 

\begin{figure}
    \includegraphics[width=.4\textwidth]{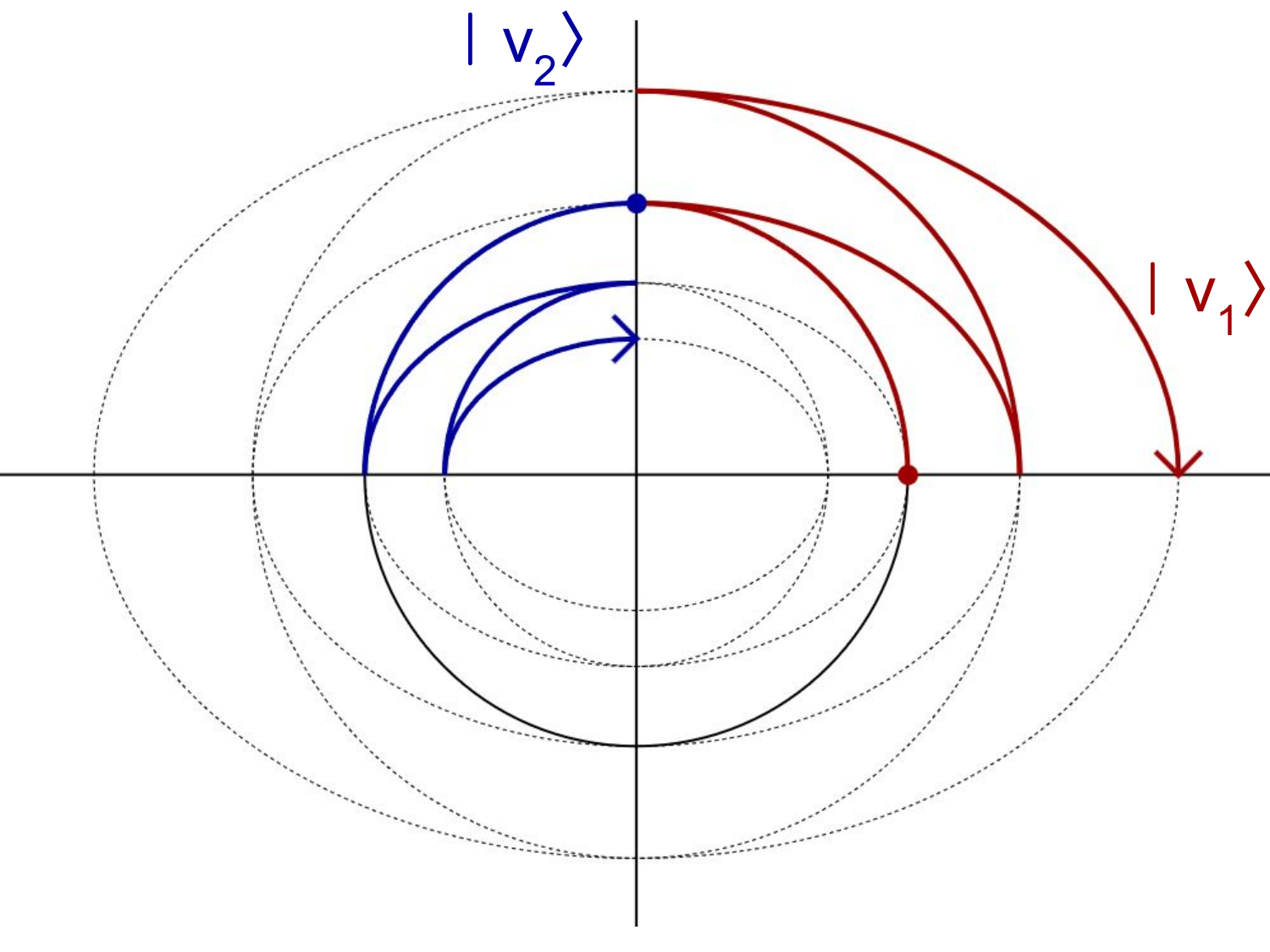}
    \vspace{-.1in}
    \caption{Geometric representation of two rounds of unitary evolution followed by $PT$-symmetric evolution for a $d=2$ system. For initial state $\ket{v_1}$ ($\ket{v_2}$), initialized along the major (minor) ellipse axis, the net effect is growth (decay) of the state norm. The unit circle is shown in solid black, while dashed lines describe the accessible circles and ellipses of other radii. When applied repeatedly to a superposition of the two states, this allows one to approximately project onto $\ket{v_1}$ up to normalization.}
    \label{fig:ellipsoid}
\end{figure}

The reason \cref{thm:NH-universality} applies even when $U=e^{-iHt}$ arises from $PT$-symmetric Hamiltonians, which exhibit quasi-periodic behavior, 
can be understood geometrically. Unitary evolution preserves the standard state norm, and so rotates states on a hypersphere in state space. Similarly, $PT$-symmetric evolution preserves the norm with respect to some metric $\eta>0$~\cite{barch2024locality, mostafazadeh_pseudo-hermiticity_2002}, and so moves states on an ellipsoid whose principal axes are set by $\eta$. 
By surrounding the $PT$-symmetric evolution by suitable unitaries (as in the SVD-based gadget), one can cancel the unitary ``rotational'' component while retaining the anisotropic stretching responsible for the postselection mechanism, as depicted in \cref{fig:ellipsoid}.

When considering gates generated by NH Hamiltonians as $U=e^{-iHt}$, the opaque nature of the SVD can make it difficult to express the required unitaries $V,W$ and repetition count $r$ directly in terms of $H$. However, for certain important subclasses, one can solve for the relevant ingredients explicitly. In particular, when $H$ is diagonalizable with complex spectrum, or when $H$ is single-qubit $PT$-symmetric, one can directly construct the required basis changes and obtain tighter, more transparent bounds on the resources needed to implement a postselection gadget. We address these cases in the next two subsections.

\subsection{Diagonalizable, Complex Spectrum}
\label{sec:diag-spectrum}

If $H$ is diagonalizable and has at least two eigenvalues with different imaginary parts, then $U=e^{-iHt}$ produces differential growth or decay between eigenspaces without the need to intersperse additional unitaries. This allows us to replace the general SVD-based discussion by an explicit diagonalization-based postselection gadget. 
Examples of such $H$ include $PT$-symmetric Hamiltonians in the broken-symmetry regime, such as the NH transverse field Ising model with strong imaginary field~\cite{barch2023scrambling, gopalakrishnan_entanglement_2021}. 
Consider the spectral decomposition
\beq
\label{eq:spec-decomp}
H = \sum_{j=1}^{d} (h_j-i\gamma_j)\ketb{r_j}{l_j},
\quad
\gamma_1\le\gamma_2\le\cdots\le\gamma_d,
\eeq
where $h_j,\gamma_j\in\mathbb{R}$, $\{\ket{r_j}\}$ and $\{\bra{l_j}\}$ are biorthonormal, and $H$ is assumed diagonalizable. Then
\beq
\label{eq:U-spec-decomp}
U = e^{-iHt} = \sum_{j=1}^{d} e^{-\gamma_j t}e^{-ih_j t}\ketb{r_j}{l_j}
\eeq
so time evolution rescales the amplitude in each right-eigenspace $j$ by $e^{-\gamma_j t}$, up to normalization.

As shown in \cref{app:deorthogonalize}, one can construct unitaries $\hat V^\dagger$ and $\hat W$ (distinct from the SVD unitaries) such that the effective map $\hat D\equiv \hat V^\dagger U \hat W$ has the two-dimensional block structure used below even when the extremal right eigenvectors $\ket{r_1}$ and $\ket{r_d}$ are non-orthogonal. 
In particular, restricted to the subspace spanned by $\ket{0^m}$ and $\ket{0^{m-1}1}$ (with $m-1$ ancilla qubits initialized to $\ket{0^{m-1}}$ as in \cref{fig:m-qubits-1-postselection}), the effective map
\begin{align}
\label{eq:diag-decay}
\begin{split}
\hat D &\equiv \hat V^\dag U \hat W\\
&= a_r e^{-\gamma_1 t}\ketbra{0^m}+a_r e^{-\gamma_d t}G+\dots\\
&\sim \ketbra{0^m}+e^{-(\gamma_d-\gamma_1)t}G+\dots
\end{split}
\end{align}
for a scalar $a_r$ and a $2\times 2$ matrix $G$ (depending on $H$ and $t$ but bounded in norm) acting on $\mathrm{span}\{\ket{0^m},\ket{0^{m-1}1}\}$.
Here, the ellipsis denotes terms that never contribute when the ancillas are initialized to $\ket{0^{m-1}}$.

Since $\gamma_d>\gamma_1$, the coefficient $e^{-(\gamma_d-\gamma_1)t}$ decays exponentially in $t$, so $\hat D$ approaches a projector onto $\ket{0^m}$ in the relevant two-dimensional block. Moreover, \cref{app:deorthogonalize} shows that $\Vert G\Vert\le \Vert G\Vert_F=1$ (where the second norm is the Frobenius norm), so this suppression bound holds uniformly, independent of the angle between $\ket{r_1}$ and $\ket{r_d}$.

To bound the induced postselection-gadget error, consider an input state $a\ket{0^m}+b\ket{0^{m-1}1}$ with $|a|^2\ge 2^{-k(n)}$ (the relevant regime for postselection). 
Using $\|G\|\le 1$, the undesired outcome probability after normalized evolution is $O\left(e^{-2(\gamma_d-\gamma_1)t}/|a|^2\right)$.
Thus, to achieve error at most $\varepsilon$, it suffices to choose $t$ such that
\beq
e^{-2(\gamma_d-\gamma_1)t}\le \varepsilon 2^{-k(n)}.
\eeq
Equivalently,
\beq
t\ge \frac{\ln(2^{k(n)}/\varepsilon)}{2(\gamma_d-\gamma_1)}.
\eeq
Taking $\varepsilon=2^{-p(n)}$ yields $t\in O((p(n)+k(n))/(\gamma_d-\gamma_1))$, which is efficient both for fixed $H$ and for $n$-dependent families $H_n$ whenever $\gamma_d-\gamma_1\in\Omega(1/\poly(n))$. Conversely, if $\gamma_d-\gamma_1$ shrinks faster than any inverse polynomial, then (potentially after an irrelevant imaginary shift) the spectrum becomes nearly real and the induced evolution is too close to unitarity to yield an efficient postselection gadget.

\subsection{Single-Qubit $PT$-Symmetric}
\label{sec:ptsym}

In the single-qubit $PT$-symmetric case (in the quasi-Hermitian sense of \cref{sec:PT-sym}), the SVD ingredients can be obtained in closed form. As shown in \cref{app:arbitrary}, after choosing an appropriate orthonormal basis and dropping terms proportional to the identity, any single-qubit $PT$-symmetric Hamiltonian can be written as
\beq
\label{eq:H-pt-sym}
H =
\begin{pmatrix}
0 & g-\gamma\\
g+\gamma & 0
\end{pmatrix},
\eeq
with $0<\gamma<g$. The corresponding time-evolution operator is
\beq
U=\cos(\omega t)\ident-\frac{i}{\omega}\sin(\omega t)H,
\quad
\omega\equiv\sqrt{g^2-\gamma^2}\in\mathbb{R}.
\eeq
At the special time $t=\pi/(2\omega)$ one has $U=-i\omega^{-1}H$. Let $\sigma^x$ denote the bit-flip operator in this basis and define
\beq
\lambda_\pm \equiv \frac{g\pm\gamma}{\omega}.
\eeq
Then
\beq
\sigma^x U
=
-i
\begin{pmatrix}
\lambda_+ & 0\\
0 & \lambda_-
\end{pmatrix},
\eeq
so $\lambda_\pm$ are the singular values of $U$ and the effective diagonal gate in the SVD gadget can be taken as $D=i\sigma^x U$ (i.e., $W=\ident$ and $V^\dag=i\sigma^x$).

To suppress the $\ket{1}$ component to error $\varepsilon=2^{-p(n)}$, it suffices to choose
\beq
r
=
\Bigl\lceil
\frac{\ln(1/\varepsilon)}{\ln(\lambda_+/\lambda_-)}
\Bigr\rceil
=
O\left(p(n)\frac{g}{\gamma}\right),
\eeq
where the final scaling follows from the lower bound
$\ln(\lambda_+/\lambda_-)=\ln\bigl((g+\gamma)/(g-\gamma)\bigr)\ge 2\gamma/g$ for $0<\gamma<g$,
so $1/\ln(\lambda_+/\lambda_-)=O(g/\gamma)$.
Thus, $r\in O(\poly(n))$ provided $\gamma/g\in\Omega(1/\poly(n))$. In this case, the only additional unitary needed to realize the postselection gadget (up to an irrelevant phase) is the bit-flip $\sigma^x$, which is proportional to the Hermitian part of $H$, namely $g\sigma^x$.

\section{Upper Bounds via Purification}
\label{sec:upperbounds}
In this section we obtain stricter upper bounds on $\mathrm{NHBQP}(U)$ and reduced non-unitary models by \emph{purifying} the non-unitary evolution into unitary evolution on a larger system together with postselection on a designated meter register.
\cref{thm:NH-universality} establishes a lower bound on worst-case hardness only in the presence of a universal family of unitary gates, so does not address non-unitary extensions of non-universal unitary models, such as Clifford circuits, matchgate circuits, or unitary dynamics efficiently simulable under tensor-networks.
Reducing non-unitary evolution to a purification gadget, composed of unitary evolution with postselection, allows us to transfer classical simulation results for the purified unitary family back to the original non-unitary dynamics.
In particular, \cref{prop:strong-closed-under-postselection} implies that strong simulability is preserved under postselection on events occurring with probability $\Omega(2^{-\poly(n)})$.
Therefore, whenever a non-unitary model admits a purification whose underlying unitary circuit family is strongly simulable, the non-unitary model itself is strongly simulable.

\subsection{Exact power of $\mathrm{NHBQP}(U)$}

We can determine the computational power of $\mathrm{NHBQP}(U)$ by combining \cref{thm:NH-universality} with a general upper bound, via a reduction to postselected unitary circuits in $\mathrm{PostBQP}$

\begin{mytheorem}[Characterization of $\mathrm{NHBQP}(U)$]
\label{thm:NHBQP-characterization}
Fix a universal finite set of unitary gates and a fixed $m$-qubit gate $U$ with $m=O(1)$ which is not proportional to a unitary, and let all gates have efficiently specifiable algebraic gate entries. Consider a uniform circuit family generated by this gate set, where each use
of $U$ is followed by renormalization. Then the class of languages decidable by this family is exactly $\mathrm{PP}$. That is,
\beq
\mathrm{NHBQP}(U) = \mathrm{PostBQP} = \mathrm{PP}.
\eeq
\end{mytheorem}

\begin{proof}
We first establish the upper bound $\mathrm{NHBQP}(U)\subseteq \mathrm{PostBQP}=\mathrm{PP}$~\cite{Aaronson:2005aa}.
The inclusion into $\mathrm{PP}$ can be obtained by extending the standard proof that $\mathrm{BQP}\subseteq\mathrm{PP}$ to the present renormalized non-unitary setting~\cite{Aaronson:2005aa, adleman1997quantum}. Here we  give an alternative proof in terms of purification into postselected unitary circuits in $\mathrm{PostBQP}$, as it motivates the ensuing simulability results for restricted systems.

Let $\{C_n\}$ be a uniform family of $\mathrm{NHBQP}(U)$ circuits deciding a language $L$ with bounded error, and assume $\{C_n\}$ is well-formed, meaning that every renormalization step is defined (equivalently, the postselected branch has nonzero norm at each such step).

We reduce $\{C_n\}$ to a postselected unitary circuit family. Since global rescalings of $U$ do not affect the induced normalized evolution, fix any $\alpha>0$ such that $K\equiv \alpha U$ satisfies $K^\dagger K\le \ident$ (e.g., $\alpha=1/\|U\|$). By the polar decomposition, $K=QR$ with $Q$ unitary and $R=(K^\dagger K)^{1/2}\ge 0$ satisfying $0\le R\le \ident$.
$R$ can be realized as a Kraus branch of a unitary $U_R$ on the system $A$ plus a single-qubit meter $B$ of the form $\bra{0}U_R\ket{0}_B = R_A$, for 
\beq
\label{eq:U_R}
    U_R =
    \left(\begin{matrix}
        R_A & \sqrt{\ident_A-R^2_A}\\
        \sqrt{\ident_A-R^2_A} & -R_A
    \end{matrix}\right)_B
\eeq
written in the computational basis of $B$.
Therefore $K$ can be implemented (up to the unitary $Q$) by a unitary circuit followed by postselection on the meter outcome.

Now replace each use of $U$ in $C_n$ by the corresponding unitary dilation gadget for $K$ and postselect the associated meter qubit on the designated outcome. Conditioned on all such postselections succeeding, the (renormalized) evolution of the system register is exactly the same as in the original $\mathrm{NHBQP}(U)$ circuit, since the overall scalar factor $\alpha$ cancels under renormalization at each step.
Because $U$ acts on $m=O(1)$ qubits, the required dilation unitaries are constant-size $m+1$ and can be approximated to inverse-exponential accuracy with only polynomial overhead using any universal unitary gate set. Finally, multiple, even interleaved, postselection conditions can be combined into a single final postselection by coherently recording each condition and postselecting on their conjunction (this does not increase the power of $\mathrm{PostBQP}$)~\cite{Aaronson:2005aa}.
Thus $L\in \mathrm{PostBQP}$, and using $\mathrm{PostBQP}=\mathrm{PP}$~\cite{Aaronson:2005aa} gives $\mathrm{NHBQP}(U)\subseteq \mathrm{PP}$.

Next we show the lower bound: $\mathrm{PP}
\subseteq \mathrm{NHBQP}(U)$.
By \cref{thm:NH-universality}, the gate $U$ together with universal unitary gates can efficiently implement a postselection gadget with error $2^{-\poly(n)}$ in the standard regime where the postselected event has probability at least $2^{-\poly(n)}$. Therefore any $\mathrm{PostBQP}$ computation can be simulated within $\mathrm{NHBQP}(U)$, 
hence we obtain $\mathrm{PP}=\mathrm{PostBQP}\subseteq \mathrm{NHBQP}(U)$.
\end{proof}

Note that we did not assume an external classical polynomial-time algorithm with adaptive oracle access to $\mathrm{NHBQP}(U)$ subroutines. Such oracle access would correspond to classes such as $\mathrm{P}^{\mathrm{PostBQP}}=\mathrm{P}^{\mathrm{PP}}=\mathrm{P}^{\#\mathrm{P}}$, which (unlike Refs.~\cite{zhang2026power,zhang2025physics}) we do not claim here.

\subsection{Purifying non-Unitary Gates into Unitary Circuits with Postselection}

\if{false}
At the gate level, one can understand the 
connection between normalized non-unitary evolution and postselection
by dilating a non-unitary map into a unitary on a larger Hilbert space followed by postselection.
Let $C$ be a fixed non-unitary gate (or circuit) acting on system $A$.
By the polar decomposition, $C=QR$ where $Q$ is unitary and $R=(C^\dagger C)^{1/2}\ge 0$.
In the normalized-evolution model, global rescalings of $C$ do not affect the induced evolution, so we may rescale $R$ by a positive constant so that $0\le R\le \ident$.
In that case $R$ can be implemented by coupling $A$ to a single-qubit meter $B$ and postselecting on $B$.
More specifically, define the unitary
\beq
\label{eq:U_R}
    U_R =
    \left(\begin{matrix}
        R_A & \sqrt{\ident_A-R^2_A}\\
        \sqrt{\ident_A-R^2_A} & -R_A
    \end{matrix}\right)_B
\eeq
written in the meter computational basis.
Then
\beq
\bra{0}U_R\ket{0}_B = R_A,
\eeq
so $C$ is realized (up to the unitary $Q$) by a unitary circuit on $AB$ with postselection on $B$.
Iterating this construction across all non-unitary steps expresses a non-unitary circuit as a unitary circuit on a larger system together with postselection on the meter outcomes.
\fi

The polar decomposition used in the proof of \cref{thm:NHBQP-characterization} is sufficient when $m=O(1)$ and one assumes access to universal unitary gates. However, it gives no bound on the resources required to implement $Q$ and $U_R$, especially when the non-unitary operator $U$ is allowed to depend on system size, and so generalizes poorly to non-universal settings. Thus, we'll instead consider the following more general construction of a purification gadget, which enables more transparent decompositions.
Consider a unitary circuit $U$ acting on a system $A$ and a meter $B$, where $B$ is initialized in a pure state $\ket{\phi}$ and then measured in the computational basis, as in \cref{fig:non-unitary-decomposition}.
Conditioning on the meter outcome $s$ induces the (unnormalized) map
\beq
\tilde{\mathcal{E}}_s(\rho) \equiv \bra{s}U\left(\rho\otimes\ketb{\phi}{\phi}\right)U^\dag\ket{s}_B,
\eeq
and the corresponding normalized (generally nonlinear) conditional state-update map
\beq
\mathcal{E}_s(\rho)\equiv \frac{\tilde{\mathcal{E}}_s(\rho)}{\Tr[\tilde{\mathcal{E}}_s(\rho)]}.
\eeq
Defining the effective (generally non-unitary) operator on $A$ by
\beq
C_s \equiv \bra{s}_B U \ket{\phi}_B,
\eeq
we have $\tilde{\mathcal{E}}_s(\rho)=C_s\rho C_s^\dag$.

\begin{figure}
\centering
\begin{quantikz}
\lstick{} & \qwbundle{n} & \gate{\mathcal{E}_s} & \qwbundle{} & \qw
\end{quantikz}
\quad = \quad
\begin{quantikz}
\lstick{\(\ket{\phi}\)} & \qwbundle{m} & \gate[wires=2]{U} & \qwbundle{} & \meter{$\ket{s}$} \qw \\
\lstick{} & \qwbundle{n} &  & \qwbundle{} & \qw
\end{quantikz}
\caption{Representation of a non-unitary (normalized) conditional map $\mathcal{E}_s$ as a purification gadget composed of a unitary circuit $U$ on a system-meter hybrid, followed by postselection of the meter measurement outcome $s$.}
\label{fig:non-unitary-decomposition}
\end{figure}

\subsection{Strong Simulability via Postselected Purification}

If the unitary circuit $U$ in a purification gadget is strongly simulable, then so are the induced postselected output probabilities (equivalently, the normalized conditional map $\mathcal{E}_s$), provided the postselection success probability is not too small.
Indeed, if one can estimate the joint probability $P_U(x,s)$ and the marginal $P_U(s)$ to exponentially small additive error, then one can compute the conditional distribution
\beq
P_{\mathcal{E}_s}(x)=\bra{x}\mathcal{E}_s(\rho)\ket{x}=\frac{P_U(x,s)}{P_U(s)}=P_U(x\mid s).
\eeq
By \cref{prop:strong-closed-under-postselection}, this computation remains efficient (in the strong-simulation sense) as long as
\beq
P_U(s)\ge 2^{-\poly(n)}.
\eeq
The same reasoning applies to conditional probabilities, e.g.,
\beq
P_{\mathcal{E}_s}(x_1\mid x_2)=P_U(x_1\mid x_2,s).
\eeq
Thus, whenever an NH model admits a purification gadget whose underlying unitary circuit family is strongly simulable, the NH model is strongly simulable as well (in the postselected/conditional sense relevant here).

Below we apply this framework to non-unitary extensions of paradigmatic strongly simulable unitary models: Clifford circuits and matchgate circuits. We then turn to tensor-network-motivated purifications that preserve locality.

\subsection{Clifford Circuits}

Clifford circuits are generated by Hadamard, phase, and CNOT gates, and form the normalizer of the Pauli group: for any Clifford circuit $U$ and Pauli string $\sigma$, $U\sigma U^\dag$ is another Pauli string.
By the Gottesman-Knill theorem, Clifford circuits with computational-basis inputs and measurements are efficiently strongly simulable~\cite{gottesman1998heisenberg, jozsa2013classical}.

As a non-unitary building block, let $U$ be an $(n+1)$-qubit Clifford circuit acting on $n$ system qubits and a one-qubit meter $B$, with $B$ initialized and postselected in $\ket{0}$.
The induced effective operation on the system is
\beq
\label{eq:post-cliff}
C \equiv \bra{0}U\ket{0}_B.
\eeq

One can also see explicitly how $C$ acts on Pauli strings.
For an $n$-qubit Pauli string $\sigma$,
\begin{align}
\begin{split}
C\sigma C^\dag
&= \bra{0}U(\sigma\otimes\ketb{0}{0})U^\dag\ket{0}_B\\
&= \frac{1}{2}\bra{0}U\bigl[\sigma\otimes(\ident+\sigma^z)\bigr]U^\dag\ket{0}_B\\
&= \frac{1}{2}\left(\bra{0}\sigma_1\ket{0}_B+\bra{0}\sigma_2\ket{0}_B\right),
\end{split}
\end{align}
where $\sigma_1\equiv U(\sigma\otimes\ident)U^\dag$ and $\sigma_2\equiv U(\sigma\otimes\sigma^z)U^\dag$ are $(n+1)$-qubit Pauli strings.
Each contraction $\bra{0}\sigma_k\ket{0}_B$ is either an $n$-qubit Pauli string (if the meter factor in $\sigma_k$ is $\ident$ or $\sigma^z$) or the zero operator (if it is $\sigma^x$ or $\sigma^y$).
Thus $C$ maps a single Pauli string to a linear combination of up to two Pauli strings, which suggests that Pauli-string tracking without meters can proliferate terms and become potentially inefficient.
Nevertheless, the overall system plus meter dilation circuit remains a Clifford circuit with a postselection, and therefore remains efficiently simulable via standard stabilizer-tableau updates conditioned on the desired measurement outcomes~\cite{Aaronson:2004aa}.
This highlights that including non-unitary non-Clifford operations (in this case $C$ arising from postselected Clifford circuits) in an otherwise Clifford circuit need not preclude efficient classical simulation. 
This contrasts the unitary case, where adding any non-Clifford \emph{unitary} to the Clifford group yields a universal gate set for $\mathrm{BQP}$ and destroys efficient classical simulation.

As a simple example, take $U=\mathrm{CNOT}_{12}$ with the meter as the target.
Then $C=\ketb{0}{0}$ is the projector onto $\ket{0}$ on the system qubit.
Circuits generated by Clifford subcircuits and such projectors were studied in Ref.~\cite{sang2023ultrafrast}, where a range of distinct conditional-information phenomena were observed, including ultrafast spreading of conditional mutual information.

Finally, note that this non-unitary effect is specific to conditioning.
If postselection is replaced by a partial trace, then for any Pauli string $\sigma$,
\begin{align}
\begin{split}
\<\sigma\>
&= \Tr\bigl[\sigma \Tr_B[U(\rho\otimes\ketb{0}{0})U^\dag]\bigr]\\
&= \Tr\bigl[(\sigma\otimes\ident_B) U(\rho\otimes\ketb{0}{0})U^\dag\bigr]\\
&= \Tr\bigl[(\rho\otimes\ketb{0}{0}) U^\dag(\sigma\otimes\ident_B)U\bigr]\\
&= \Tr\bigl[\rho \expval{0}{U^\dag(\sigma\otimes\ident_B)U}_B\bigr],
\end{split}
\end{align}
so the unconditional channel does not introduce the same potential for Pauli-string branching.

\subsection{Matchgate Circuits}

Matchgate circuits form another paradigmatic strongly simulable family~\cite{valiant, jozsa2008matchgates, shtanko2021complexity, projansky2025gaussianity}.
They are generated by two-qubit unitary gates of the form
\begin{align}
    U(F,G) = \left(\begin{matrix}
        F_{11} & 0 & 0 & F_{12}\\
        0 & G_{11} & G_{12} & 0\\
        0 & G_{21} & G_{22} & 0\\
        F_{21} & 0 & 0 & F_{22}
    \end{matrix}\right),
\end{align}
with $\det(F)=\det(G)$.
Circuits composed of nearest-neighbor matchgates on a 1D chain, with product-state inputs and computational-basis measurements, can be strongly simulated efficiently via a mapping to free fermions (Jordan-Wigner transformation).

A simple non-unitary extension is obtained by coupling a system qubit to a one-qubit meter via a matchgate, then postselecting the meter.
When the meter is initialized to $\ket{+}$ and postselected to $\ket{0}$, the induced single-qubit map on the system is
\beq
\label{eq:matchgate-induced}
\bra{0}U(F,G)\ket{+}_B
=
\frac{1}{\sqrt{2}}
\left(\begin{matrix}
        F_{11} & F_{12}\\
        G_{21} & G_{22}
\end{matrix}\right).
\eeq
The choice of $F$ and $G$ subject to matchgate constraints allows this to produce a broad family of single-qubit system operators, including arbitrary unitary gates under normalized evolution.
The matchgate determinant condition $\det(F)=\det(G)$ can always be satisfied by adjusting phases on the unused rows of $F$ and $G$, and it does not restrict the induced single-qubit map in \cref{eq:matchgate-induced}.

Our purification-based simulation applies only when the enlarged unitary circuit (system plus meters) remains within the standard strongly simulable class of 1D nearest-neighbor matchgate circuits. Since each effective gate of the form in \cref{eq:matchgate-induced} requires coupling to a unique meter qubit, such gates cannot be inserted arbitrarily along the chain without leaving this class. In particular, to preserve a 1D nearest-neighbor geometry, they can be placed only at the ends of the chain, so at most two such gates can be included. This amounts to only a constant overhead even under non-purified simulation. This restriction is in fact essential: if gates of the form in \cref{eq:matchgate-induced} could be applied freely as primitives in a matchgate circuit, then, since they realize arbitrary single-qubit unitaries under normalization, the resulting model would be universal for $\mathrm{BQP}$. We do not attempt to characterize the most general class of postselected matchgate circuits, but simply take this case as an example.

\subsection{Trotterized NH Dynamics}

The Trotterization framework used to simulate Lindblad evolution~\cite{cleve2017efficient, ding2024simulating} can be adapted to conditional (trajectory) evolution by conditioning on specific meter outcomes (postselection), rather than averaging over them (partial trace).
Operationally, one alternates short unitary interactions between a system and a meter with measurements of the meter (\cref{fig:non-unitary-timestep}).
Averaging over meter outcomes yields unconditional (Lindbladian) dynamics, while 
conditioning on the meter remaining in its initial state at every step produces the no-jump trajectory and evolution under an effective NH Hamiltonian on the system.

The construction below is a discrete-time analogue of the standard no-jump derivation~\cite{ashida2020nonhermitian}, formulated in such a way that the system-meter unitary can be chosen to preserve locality, as required by tensor-network simulations.
A derivation that treats general trajectories including jump events, and the unconditional Lindblad limit, is given in \cref{app:trajectories-purified}.

Fix a meter basis $\{\ket{j}\}$ and consider a system $A$ coupled to a meter $B$ with Hamiltonian
\beq
H \equiv H_A + \frac{1}{\sqrt{\delta}} H_{AB},
\eeq
where $\delta>0$ is the timestep, and
\beq
\label{eq:H-meter}
H_A \equiv \sum_j H_j\otimes\ketb{j}{j},
\quad
H_{AB} \equiv \sum_{j\neq k} L_{jk}\otimes\ketb{j}{k}.
\eeq
Hermiticity of $H_{AB}$ implies $L_{jk}=L_{kj}^\dagger$, and the scaling $H_{AB}/\sqrt{\delta}$ is the standard choice that makes jump amplitudes $O(\sqrt{\delta})$ and jump probabilities $O(\delta)$ per timestep.

For a single step, let
\beq
\label{eq:U-timestep}
U \equiv e^{-iH\delta} = e^{-i(\delta H_A + \sqrt{\delta} H_{AB})}.
\eeq
Expanding in $\sqrt{\delta}$ gives
\beq
U
=
\ident
-i\sqrt{\delta}H_{AB}
-i\delta H_A
-\frac{\delta}{2}H_{AB}^2
+O(\delta^{3/2}).
\eeq
The $O(\sqrt{\delta})$ term is off-block-diagonal in the meter basis, since $H_{AB}$ maps $\ket{j}$ to $\ket{k\neq j}$.

\subsubsection{No-jump (drift) timestep}

Initialize the meter in $\ket{0}$ and postselect the measurement outcome $\ket{0}$ at the end of the step, as in \cref{fig:non-unitary-timestep}.
The induced (unnormalized) system map is the Kraus operator
\beq
\label{eq:drift-evolution}
C \equiv \expval{0}{U}_B.
\eeq
Using the expansion above, the $\sqrt{\delta}$ term vanishes under $\expval{0}{\cdot}_B$, and one obtains
\beq
\label{eq:drift-C}
C
=
\ident_A
-i\delta \expval{0}{H_A}_B
-\frac{\delta}{2}\expval{0}{H_{AB}^2}_B
+O(\delta^{3/2}).
\eeq

Writing Hermitian system Hamiltonian $\expval{0}{H_A}_B=H_0$ and using
\beq
\expval{0}{H_{AB}^2}_B = \sum_{j\neq 0} L_{0j}L_{j0} = \sum_{j\neq 0} L_{j0}^\dag L_{j0},
\eeq
we obtain
\beq
\begin{aligned}
C
&=
\ident_A
-i\delta\left(H_0-\frac{i}{2}\sum_{j\neq 0} L_{j0}^\dag L_{j0}\right)
+O(\delta^{3/2})\\
&=
e^{-i\delta H_\mathrm{eff}^0}+O(\delta^{3/2}),
\end{aligned}
\eeq
where the no-jump effective Hamiltonian is
\beq
\label{eq:H_0_eff}
\begin{aligned}
H_\mathrm{eff}^0
&\equiv
H_0-\frac{i}{2}\sum_{j\neq 0} L_{j0}^\dag L_{j0}\\
&=
\expval{0}{H_A}_B-\frac{i}{2}\expval{0}{H_{AB}^2}_B.
\end{aligned}
\eeq
Up to an overall imaginary shift (physically irrelevant under normalized evolution), this construction can generate an arbitrary NH Hamiltonian on $A$, since $\sum_j L_{j0}^\dag L_{j0}\ge 0$ but is otherwise unconstrained.

The Trotter error can be reduced from $O(\delta^{3/2})$ to $O(\delta^2)$ by picking $H_{AB}$ to only include $L_{jk}$ terms with $j=0$ or $k=0$. Under this choice, odd powers of $H_{AB}$ are off-diagonal in the meter basis, causing the $O(\delta^{3/2})$ term to drop out of \cref{eq:drift-C}. While this choice would not affect $H_\mathrm{eff}^0$, it would complicate the construction of other $H_\mathrm{eff}^k$ for $k\neq0$.

\begin{figure}
\centering
\begin{quantikz}
\lstick{\(\ket{0}\)} & \qwbundle{} & \gate[wires=2]{e^{-iH_{AB}\sqrt{\delta}}} & \qwbundle{} & \meter{$\ket{0}$} \qw \\
\lstick{} & \gate{e^{-iH_A\delta}} &  & \qwbundle{} & \qw
\end{quantikz}
\caption{A single timestep of the purification gadget implementing no-jump NH evolution.
The unitary $e^{-iH_A\delta}$ is block-diagonal in the meter basis (a controlled evolution), while $e^{-iH_{AB}\sqrt{\delta}}$ generates $O(\sqrt{\delta})$ transitions in the meter.
Postselecting on the meter reproduces the no-jump Kraus operator $C=\expval{0}{U}_B$ to $O(\delta)$, hence $H_\mathrm{eff}^0$ to leading order.}
\label{fig:non-unitary-timestep}
\end{figure}

\subsubsection{Other no-jump sectors and jump events}

If instead the meter is initialized and postselected in $\ket{k}$, the same derivation yields a (generally different) no-jump Hamiltonian
\beq
\label{eq:H_k_eff}
H_\mathrm{eff}^k
\equiv
H_k-\frac{i}{2}\sum_{j\neq k} L_{jk}^\dag L_{jk}.
\eeq
Additionally, if the meter begins in $\ket{j}$ and is postselected in $\ket{k\neq j}$, the leading contribution comes from the $O(\sqrt{\delta})$ part of $U$ and produces a jump map.
As shown in \cref{app:trajectories-purified}, the induced unnormalized update is
\beq
\label{eq:jump}
\tilde{\mathcal{E}}_{k\leftarrow j}(\rho)
=
\delta L_{kj}\rho L_{kj}^\dag
+O(\delta^{3/2}),
\eeq
which becomes the corresponding quantum-jump update after renormalization.
Together, the family of no-jump maps and jump maps realizes a discrete-time quantum trajectories model specified by $\{H_k\}_k$ and $\{L_{kj}\}_{k\neq j}$.

\subsubsection{A locality-preserving choice}

A key advantage of this dynamical purification gadget is that $H$ can be chosen so that locality in the system is preserved by the system-meter unitary $U$, and hence by tensor-network representations thereof.
Suppose the desired jump operators are local, with $L_{j0}=L_{j_A}$ acting on a localized subregion $j_A$ of $A$.
Introduce a meter $B$ with local subregions $j_B$ paired to $j_A$, and define
\beq
\label{eq:local-H}
H_A \equiv H_0\otimes\ident_B,
\quad
H_{AB} \equiv \sum_j \left(L_{j_A}\otimes\sigma^+_{j_B} + L_{j_A}^\dag\otimes\sigma^-_{j_B}\right).
\eeq
Taking the meter input and postselection state to be $\ket{0}_B\equiv\ket{0^m}_B$, we obtain
\beq
\begin{aligned}
\label{eq:trotter-MPO}
H_\mathrm{eff}^0
&=
H_0-\frac{i}{2}\sum_{jk} L_{j_A}^\dag L_{k_A}\expval{0^m}{\sigma^-_{j_B}\sigma^+_{k_B}}_B\\
&=
H_0-\frac{i}{2}\sum_j L_{j_A}^\dag L_{j_A}.
\end{aligned}
\eeq
Because the expectation $\expval{0^m}{\sigma^-_{j_B}\sigma^+_{k_B}}_B$ vanishes unless $j=k$, the anti-Hermitian correction is a sum of local terms $L_{j_A}^\dag L_{j_A}$.
In particular, the interaction graph induced by $H_\mathrm{eff}^0$ is determined by $H_0$ together with said local terms, so it preserves the same notion of locality as the underlying unitary system dynamics.
A representative 1D example is shown in \cref{fig:trotterMPO}.

This same construction yields the jump terms $\{L_{j_A}\rho L_{j_A}^\dag\}_j$ by postselecting the corresponding meter excitations, and yields the full Lindblad evolution if meter outcomes are traced rather than postselected, as reviewed in \cref{app:trajectories-purified}.
Multiple jump operators per site can be included by taking $j_B$ to be a small register and coupling each local operator to a distinct raising operator within that register.

When extending from a single system Hamiltonian $H_0$ to a family $\{H_k\}_k$ controlled by meter outcomes, locality requires that changes in the meter state only induce local changes in the system Hamiltonian.
One sufficient condition is that $H_A$ decomposes into local controlled terms,
\beq
H_A = \sum_j H_{j_{AB}},
\quad
H_{j_{AB}} \equiv \sum_i h_{j_A}^i\otimes\ketbra{i}_{j_B},
\eeq
where each $h_{j_A}^i$ acts only on a neighborhood of $A$ around $j_A$.
More general trajectory models can also be implemented by allowing feedback between steps (choosing the next $H$ conditioned on measured meter outcomes), at the cost of stronger control assumptions.

Finally, this locality-preserving purification gadget construction connects directly to simulability.
If the unitary system-meter circuit $U$ is strongly simulable (for example, via an exact or exponentially accurate tensor-network contraction), then by \cref{prop:strong-closed-under-postselection} the induced conditional dynamics are also strongly simulable, provided the postselection success probabilities remain at least $2^{-\poly(n)}$ for the trajectories of interest.

\begin{figure}
\centering
\begin{tikzpicture}
    \foreach \i in {1,2,3} {
        \node[circle,draw,minimum size=6mm,inner sep=0pt] (\i A) at ({(\i-1)*2},0) {\(\i_A\)};
        \node[circle, draw,minimum size=6mm,inner sep=0pt] (\i B) at ({(\i-1)*2},2.5) {\(\i_B\)};
    }
    \draw[blue, thick] (1A) -- (2A) node[midway, below=2pt, blue] {\( h_{1_A 2_A} \)};
    \draw[blue, thick] (2A) -- (3A) node[midway, below=2pt, blue] {\( h_{2_A 3_A} \)};
    \draw[blue, thick] ($(1A) + (-1,0)$) -- (1A);
    \draw[blue, thick] (3A) -- ($(3A) + (1,0)$);

    \draw[red, thick] (1A) -- (1B);
    \draw[red, thick] (2A) -- (2B);
    \draw[red, thick] (3A) -- (3B);

    \node[red] at ($(1A)!0.5!(1B) + (-0.8,0.1)$) {\( L_{1_A} \sigma_{1_B}^{+} \)};
    \node[red] at ($(1A)!0.5!(1B) + (-0.7,-0.4)$) {+ h.c.};

    \node[red] at ($(2A)!0.5!(2B) + (-0.8,0.1)$) {\( L_{2_A} \sigma_{2_B}^{+} \)};
    \node[red] at ($(2A)!0.5!(2B) + (-0.7,-0.4)$) {+ h.c.};

    \node[red] at ($(3A)!0.5!(3B) + (-0.8,0.1)$) {\( L_{3_A} \sigma_{3_B}^{+} \)};
    \node[red] at ($(3A)!0.5!(3B) + (-0.7,-0.4)$) {+ h.c.};
\end{tikzpicture}
\caption{Locality-preserving couplings in \cref{eq:local-H} for a 1D chain system Hamiltonian $H_0=\sum_{\langle i,j\rangle} h_{ij}$ on $A$ with single-qubit $L_{j_A}$. The system-system couplings (blue) and system-meter couplings (red) together define a local system-meter Hamiltonian $H$ whose no-jump sector yields $H_\mathrm{eff}^0$ in \cref{eq:trotter-MPO}.}
\label{fig:trotterMPO}
\end{figure}

\subsection{$PT$-Symmetric Dynamics}

When the NH Hamiltonian $H$ is quasi-Hermitian (equivalently, $PT$-symmetric in the unbroken phase), questions about $U=e^{-iHt}$ can be reduced to the corresponding questions about a unitary evolution.
As discussed in \cref{sec:PT-sym}, in this regime there exists a positive definite metric $\eta>0$ such that $H^\dagger\eta=\eta H$.\footnote{In the broken $PT$-symmetric phase no such $\eta>0$ exists, and the reduction below does not apply.}
Defining $S \equiv \eta^{-1/2}>0$,
one obtains a Hermitian Hamiltonian
$H_0 \equiv S^{-1} H S$,
and hence the similarity transform
$H = SH_0S^{-1}$, with $U = SU_0 S^{-1}$ and $U_0\equiv e^{-iH_0t}$.

If $U_0$ admits an efficient tensor-network representation (and contraction algorithm) with bond dimension $f(n)$, and if both $S$ and $S^{-1}$ admit MPO representations of bond dimension $q(n)\in O(\poly(n))$, then $U=SU_0S^{-1}$ has an MPO representation of bond dimension at most $f(n)q^2(n)$.
Thus, in this regime, tensor-network simulability of $U$ follows from that of $U_0$ with only polynomial overhead. An example of such a system is the 1D NH Transverse-Field Ising Model~\cite{gopalakrishnan_entanglement_2021, barch2023scrambling}, for which $S$ is a product operator.
For general $S$ the required bond dimension may be exponential, in which case the map $U=SU_0S^{-1}$ need not preserve efficient tensor-network simulability.

It is also useful to represent the similarity transform as a purification gadget~\cite{karuvade2022observing}.
The one-qubit purification in \cref{eq:U_R} applies to any positive operator that is a contraction.
Accordingly, pick constants $\alpha,\beta>0$ such that $\alpha S\le \ident$ and $\beta S^{-1}\le \ident$.
Then there exist unitaries $V_S$ 
on $A\otimes B$ and $V_{S^{-1}}$ on $A\otimes B'$ (with single-qubit meters $B,B'$) such that
\beq
\expval{0}{V_S}_B = \alpha S,
\quad
\expval{0}{V_{S^{-1}}}_{B'} = \beta S^{-1}.
\eeq
With two meter qubits initialized and postselected in $\ket{0}$  as in \cref{fig:PT-Sym-circuit}, the effective operator is
\beq
\expval{00}{V_{S} U_0 V_{S^{-1}}}_{BB'}
=
\alpha\beta SU_0S^{-1}
=
\alpha\beta U,
\eeq
which equals $U$ up to an overall scalar.
This scalar does not affect normalized evolution or strong-simulation statements, but it does affect physical success probabilities of postselection.

\begin{figure}[t]
\centering
\begin{quantikz}
\lstick{} & \qwbundle{n} & \gate{U} & \qwbundle{}
\end{quantikz}
\quad = \quad
\begin{quantikz}
\lstick{\(\ket{0}\)} & \qw & \gate[wires=2]{V_{S^{-1}}} & \qw & \qw & \meter{\(\ket{0}\)} \qw \\
\lstick{} & \qwbundle{n} &  & \gate{U_0} & \gate[wires=2]{V_{S}} & \qwbundle{} \\
\lstick{\(\ket{0}\)} & \qw & \qw & \qw & \qw & \meter{\(\ket{0}\)} \qw
\end{quantikz}
\caption{Two-meter purification gadget for $U=SU_0S^{-1}$ for $PT$-symmetric dynamics.
The unitaries $V_S$ and $V_{S^{-1}}$ satisfy $\expval{0}{V_S}=\alpha S$ and $\expval{0}{V_{S^{-1}}}=\beta S^{-1}$ for suitable scalars $\alpha,\beta>0$ that make the implemented operators contractions.
Postselecting both meters in $\ket{0}$ yields $(\alpha\beta)U$.}
\label{fig:PT-Sym-circuit}
\end{figure}

When $S$ factorizes across local subregions, the purification gadget preserves locality in a particularly transparent way.
For example, if $S=\bigotimes_j S_{j_A}$ is a product operator on local subregions $j_A$ of $A$, then one can choose $V_S$ as a product of two-qubit gates $V_S=\bigotimes_j V_{j_A j_B}$ coupling each $j_A$ to a local meter qubit $j_B$, with
\beq
\expval{0}{V_{j_A j_B}}_{j_B} = \alpha_j S_{j_A}
\eeq
for appropriate local rescalings $\alpha_j$.
In this case both $V_S$ and $V_{S^{-1}}$ can be expressed as unitary MPOs of constant bond dimension, and the unitary dilation introduces only local system-meter couplings, as depicted in \cref{fig:product-Vs}.

\begin{figure}
\centering
\begin{tikzpicture}
    \foreach \i in {1,2,3} {
        \node[circle,draw,minimum size=6mm,inner sep=0pt] (\i A) at ({(\i-1)*2},0) {\(\i_A\)};
        \node[circle, draw,minimum size=6mm,inner sep=0pt] (\i B) at ({(\i-1)*2},2.5) {\(\i_B\)};
    }
    \draw[red, thick] (1A) -- (1B);
    \draw[red, thick] (2A) -- (2B);
    \draw[red, thick] (3A) -- (3B);

    \node[red] at ($(1A)!0.5!(1B) + (-0.8,0.1)$) {\( V_{1_A 1_B} \)};
    \node[red] at ($(2A)!0.5!(2B) + (-0.8,0.1)$) {\( V_{2_A 2_B} \)};
    \node[red] at ($(3A)!0.5!(3B) + (-0.8,0.1)$) {\( V_{3_A 3_B} \)};
\end{tikzpicture}
\caption{Product structure of the unitary $V_S$ when $S$ is a product operator, yielding a locality-preserving purification gadget.}
\label{fig:product-Vs}
\end{figure}

\section{Discussion}
\label{sec:conclusion}

This work establishes a bidirectional relationship between normalized non-Hermitian (NH) dynamics and postselection.
On one hand, \cref{sec:NHBQP} approximates an explicit postselection gadget to exponentially small error, built from normalized evolution under a coherent non-unitary gate together with suitable surrounding unitaries (in our construction, $V^\dagger$ and $W$ arising from an SVD of the non-unitary gate).
For fixed-support gates this applies to any gate not proportional to a unitary, and for size-dependent gate families it applies under the explicit uniform implementability and inverse-polynomial singular-value-gap assumptions stated there.
This leads to our central complexity-theoretic implication (\cref{thm:NH-universality}): universal unitary gates supplemented with any fixed gate $U$ that is not proportional to a unitary, together with renormalization after each use of $U$ (equivalently, renormalizing the output probabilities during measurement), can implement a postselection gadget and therefore decide every language in $\mathrm{PostBQP}$.
Moreover, \cref{thm:NHBQP-characterization} shows this is tight in the uniform circuit-family model: $\mathrm{NHBQP}(U)=\mathrm{PostBQP}=\mathrm{PP}$ for any fixed such gate $U$.

On the other hand, \cref{sec:upperbounds} shows that a broad class of non-unitary evolutions can be \emph{purified} into unitary evolution on a larger system followed by postselection on a meter register, thereby inheriting strong simulability from the unitary supersystem where applicable. 
As a special case, a single time step of a general quantum-trajectory update can be realized as unitary evolution on a larger system followed by postselection on a meter register, with a per-timestep error $O(\delta^{3/2})$ for timestep $\delta$.

A key message is that the above reductions identify \emph{sufficient} conditions for equivalence, but they are not obviously \emph{necessary}.
For example, the SVD-based postselection gadget in \cref{sec:NHBQP} makes the role of a singular-value gap explicit, but it hides the dependence of the surrounding unitaries on microscopic descriptions such as a local Hamiltonian $H$, which could allow one to better bound the cost of implementing them.
Likewise, the purification gadget constructions provide a route to simulation and upper bounds, but their quantitative performance is tied to the chosen dilation and to discretization error (e.g., for the Trotterized approach, $O(\delta^{3/2})$ per step).
It would be valuable to develop postselection gadgets that can be implemented more directly from a local NH generator, and to sharpen the resource analysis (gate complexity, success probability, and accumulated error over long times) for physically motivated families of NH systems.

Our simulability results in \cref{sec:BG} and \cref{sec:upperbounds} are inherently \emph{strong}-simulation statements, in that they rely on the ability to estimate relevant probabilities to exponentially small additive error, which is what is needed to remain stable under conditioning.
If the purified unitary model is, for example, only \emph{weakly} simulable (sampling without probability estimation), then \cref{prop:strong-closed-under-postselection} does not apply, and postselection can in general increase classical complexity substantially.
Understanding when sub-strong-simulation results can be transferred through purification, and what sampling complexity classes are natural for restricted NH models, remains an important open direction.

Although NH evolution is unlikely to be efficiently scalable as a computational primitive in the universal-unitary setting suggested by \cref{thm:NH-universality}, NH physics has already been realized experimentally at modest sizes, particularly in photonics where postselection and loss engineering are natural ingredients.
Relatedly, postselection on ``no error detected'' is implicit in successful runs of error-detection protocols, so the corresponding conditional evolution is effectively non-unitary.
Even if such procedures do not scale efficiently in the worst case, they can serve as platforms for probing NH phenomena and for demonstrating postselection-gadget behavior at small sizes in settings where conditioning is already part of the experimental setup.
It is also interesting to explore proposals in which NH dynamics might emerge without explicit postselection, for example via  special relativity as discussed in Ref.~\cite{Paiva_2022}. If physically realized in a controllable way, such effects would have striking complexity-theoretic implications.

Finally, the no-jump trajectory generating NH systems makes up a small subset of all possible conditional evolutions.
More general conditional dynamics are described by quantum trajectories: NH drift interspersed with stochastic jumps.
It remains an open question how robust NH-induced dynamical signatures (such as locality breakdown) are under a limited number of jumps. For example, Ref.~\cite{ashida2018full} gives evidence for persistent Lieb-Robinson bound violations with a strength that decays rapidly with the number of jumps.
While the ensemble-averaged evolution is linear and CPTP, the state conditioned on an observed measurement record is trace-normalized and therefore evolves via a nonlinear Bayesian update rule.
However, in the standard circuit model, stochastic trajectories together with classical feedforward are already encompassed by $\mathrm{BQP}$; achieving postselection-like computational power would require conditioning on \emph{particular} (potentially exponentially unlikely) records, i.e., explicit postselection.
An interesting open direction is to characterize the complexity of restricted measurement-and-feedback/trajectory models and to quantify when, and at what cost in success probability or resources, trajectory-level conditioning can emulate the postselection gadgets studied here.

\acknowledgments
This material is based upon work supported by, or in part by, the U. S. Army Research Laboratory and the U.S. Army Research Office under contract/grant number W911NF2310255. 
We are grateful to Qi Zhang and Biao Wu for insightful discussions about computational complexity (see \cref{app:ZhangWu} for a discussion of the relation between their work~\cite{zhang2025physics} and ours).

\appendix
\begin{appendix}
\numberwithin{equation}{section}

\section{Proofs from Background}
\subsection{Pseudo-Hermitian $H$ has Indefinite $\Gamma$}
\label{app:PT-indefinite-gamma}

Here we prove the statement in \cref{sec:BG-NH} that any diagonalizable pseudo-Hermitian Hamiltonian has indefinite anti-Hermitian part.

\begin{proof}
Suppose $H=H_0-i\Gamma$ with $H_0=H_0^\dagger$ and $\Gamma=\Gamma^\dagger\ge 0$, and let $H\ket{r_j}=\nu_j\ket{r_j}$.
Normalizing $\ket{r_j}$ so that $\bk{r_j}{r_j}=1$ gives
\beq
\nu_j=\bra{r_j}H\ket{r_j}=\bra{r_j}H_0\ket{r_j}-i\bra{r_j}\Gamma\ket{r_j},
\eeq
so $\Im(\nu_j)=-\bra{r_j}\Gamma\ket{r_j}\le 0$.
Pseudo-Hermiticity implies the spectrum is closed under complex conjugation: if $\nu$ is an eigenvalue then $\nu^\ast$ is also an eigenvalue. Together with $\Im(\nu_j)\le 0$ for all $j$, this forces $\Im(\nu_j)=0$ for all $j$ (otherwise the conjugate eigenvalue would have positive imaginary part). Hence every $\nu_j$ is real and thus $\bra{r_j}\Gamma\ket{r_j}=0$ for all $j$.
Since $\Gamma\ge 0$, this forces $\Gamma\ket{r_j}=0$ for all $j$.
If $H$ is diagonalizable, the right eigenvectors span the space, so $\Gamma=0$.
\end{proof}

\subsection{Closure of Strong Simulability under Postselection}
\label{app:strong-closed-under-postselection}

Here we provide the proof of \cref{prop:strong-closed-under-postselection}, showing closure of strong simulability under postselection.

\begin{proof}
Fix $n$ and abbreviate
\beq
A \equiv P_n(x,s^\star),
\quad
B \equiv P_n(s^\star),
\eeq
so that $0\le A\le B$ and $P_n^{\mathrm{post}}(x)=A/B$.
Let $\tilde A$ and $\tilde B$ be estimates of $A$ and $B$ with additive error at most $\epsilon'$:
\beq
|\tilde A-A|\le \epsilon',
\quad
|\tilde B-B|\le \epsilon'.
\eeq
Strong simulability of $P_n(x,s^\star)$ implies that $\tilde A$ and $\tilde B$ can be classically calculated in time $\poly(n,\log(1/\epsilon'))$. Assume $\epsilon'\le B/2$.
Then
\beq
\tilde B \ge B-\epsilon' \ge B/2>0,
\eeq
so the denominator $\tilde B$ is strictly positive.
Define the estimate
\beq
\widetilde{P}_n^{\mathrm{post}}(x)\equiv \frac{\tilde A}{\tilde B}.
\eeq
Its error obeys
\begin{align}
\begin{split}
\left|\frac{\tilde A}{\tilde B}-\frac{A}{B}\right|
&=
\left|\frac{\tilde A B-A\tilde B}{\tilde B B}\right|
=
\left|\frac{B(\tilde A-A)+A(B-\tilde B)}{\tilde B B}\right|\\
&\le
\frac{B|\tilde A-A|+A|B-\tilde B|}{|\tilde B| B}
\le
\frac{B\epsilon'+A\epsilon'}{(B/2)B}
\le \frac{4\epsilon'}{B}.
\end{split}
\end{align}

Therefore, to guarantee conditional additive error at most $\epsilon$, we may assume without loss of generality that $0<\epsilon\le 1$ (otherwise set $\epsilon\equiv 1$), and choose
\beq
\epsilon' \equiv \frac{\epsilon}{4}2^{-q(n)}.
\eeq
Since $B\ge 2^{-q(n)}$, this choice ensures both $\epsilon'\le (\epsilon/4)B\le B/4$ (hence $\epsilon'\le B/2$) and
\beq
\frac{4\epsilon'}{B}\le \epsilon,
\eeq
so the ratio estimate satisfies $\bigl|\widetilde{P}_n^{\mathrm{post}}(x)-P_n^{\mathrm{post}}(x)\bigr|\le \epsilon$.

Finally, with this choice,
\beq
\log(1/\epsilon')=\log(1/\epsilon)+O(q(n)),
\eeq
and since $q(n)$ is polynomial in $n$, a runtime $\poly\bigl(n,\log(1/\epsilon')\bigr)$ is also $\poly\bigl(n,\log(1/\epsilon)\bigr)$.
\end{proof}

\section{Distance from Unitarity}
\label{app:unitary-distance}
Let $U\in\mathbb{C}^{d\times d}$ be nonzero and not proportional to a unitary, with singular values
$\lambda_1\ge \cdots \ge \lambda_d\ge 0$ (so $\lambda_1>0$),
and recall the normalized singular radius $\Delta \equiv 1-\frac{\lambda_d}{\lambda_1}\in(0,1]$.
Throughout this appendix, $\|\cdot\|$ denotes the operator norm, i.e. the largest singular value.

\begin{mylemma}[Distance to the unitary group up to rescaling]
\label{lem:unitary-distance}
Let $U=VDW^\dagger$ be a singular value decomposition, with $V,W$ unitary and
$D=\mathrm{diag}(\lambda_1,\ldots,\lambda_d)\ge 0$.
Then
\beq
\begin{aligned}
\min_{\alpha>0, T\in\mathcal{U}(d)} \|\alpha U-T\|
&=
\frac{\lambda_1-\lambda_d}{\lambda_1+\lambda_d}
=
c \Delta \\
c&\equiv \frac{\lambda_1}{\lambda_1+\lambda_d}\in(\frac{1}{2},1].
\end{aligned}
\eeq
Equivalently,
\beq
\Delta = \frac{1}{c}\min_{\alpha>0, T\in\mathcal{U}(d)} \|\alpha U-T\|.
\eeq
\end{mylemma}

\begin{proof}
Fix $\alpha>0$. By Fan's theorem~\cite{fan1955some}, among all unitaries $T$ the minimizer of
$\|\alpha U-T\|$ is the polar factor of $\alpha U$, which for an SVD is $VW^\dagger$.
Thus
\beq
\min_{T\in\mathcal{U}(d)} \|\alpha U-T\| = \|\alpha U - VW^\dagger\|.
\eeq
Using unitary invariance of the operator norm,
\beq
\|\alpha U - VW^\dagger\|
=
\|V(\alpha D-\ident)W^\dagger\|
=
\|\alpha D-\ident\|.
\eeq
Since $\alpha D-\ident$ is diagonal, and since $\lambda_d\le \lambda_j\le \lambda_1$,
\beq
\|\alpha D-\ident\|
=
\max_{j} |\alpha\lambda_j-1|
=
\max\left\{|\alpha\lambda_1-1|, |\alpha\lambda_d-1|\right\}.
\eeq

The function
$f(\alpha)=\max\{|\alpha\lambda_1-1|,|\alpha\lambda_d-1|\}$
is minimized when $\alpha\lambda_1-1 = 1-\alpha\lambda_d$, so the minimizing value of $\alpha$ is
\beq
\alpha^\star = \frac{2}{\lambda_1+\lambda_d}.
\eeq
At this value,
\beq
f(\alpha^\star)
=
\left|\frac{2\lambda_1}{\lambda_1+\lambda_d}-1\right|
=
\frac{\lambda_1-\lambda_d}{\lambda_1+\lambda_d}.
\eeq
This proves the claimed formula and the identification with $c \Delta$.
\end{proof}

\section{Constructing the unitaries $\hat V,\hat W$ in
\cref{eq:diag-decay}}
\label{app:deorthogonalize}

Here we show how to construct unitaries $\hat V^\dagger$ and $\hat W$ such that
$\hat V^\dagger U \hat W$ has the two-dimensional block structure used in
\cref{eq:diag-decay}, for a diagonalizable NH evolution with complex spectrum.

Let $H$ be diagonalizable with right and left eigenvectors $\{\ket{r_j}\}$ and $\{\bra{l_j}\}$,
biorthonormalized so that $\brak{l_i}{r_j}=\delta_{ij}$.
Write $H$ as in \cref{eq:spec-decomp},
and normalize the right eigenvectors so that $\brak{r_j}{r_j}=1$ for all $j$.
Then the time evolution operator $U = e^{-iHt}$ is given by \cref{eq:U-spec-decomp}.

Fix right eigenvectors $\ket{r_1}$ and $\ket{r_d}$ corresponding to extremal decay rates
$\gamma_1$ and $\gamma_d$ (as in \cref{eq:spec-decomp}), 
and define their overlap
$b_r \equiv \brak{r_d}{r_1}$.
Since $\ket{r_1}$ and $\ket{r_d}$ are linearly independent and normalized, $|b_r|<1$.
Define normalized vectors in $\mathrm{span}\{\ket{r_1},\ket{r_d}\}$ orthogonal to
$\ket{r_d}$ and $\ket{r_1}$, respectively:
\beq
\ket{r_1^\perp} \equiv a_r\left(\ket{r_1}-b_r\ket{r_d}\right),
\quad
\ket{r_d^\perp} \equiv a_r\left(\ket{r_d}-b_r^*\ket{r_1}\right),
\eeq
where $a_r\equiv \frac{1}{\sqrt{1-|b_r|^2}}$. Then $\{\ket{r_1^\perp},\ket{r_d}\}$ and $\{\ket{r_1},\ket{r_d^\perp}\}$ are orthonormal pairs and
can be completed to orthonormal bases of the full Hilbert space.

Using these pairs, define unitaries $\hat W$ and $\hat V^\dagger$ by their action on the two
computational basis states of interest:
\beq
\begin{aligned}
\hat W
&\equiv
e^{i h_1 t}\ketb{r_1^\perp}{0^m}
+
e^{i h_d t}\ketb{r_d}{0^{m-1}1}
+\cdots\\
\hat V^\dagger
&\equiv
\ketb{0^m}{r_1}
+
\ketb{0^{m-1}1}{r_d^\perp}
+\cdots,
\end{aligned}
\eeq
where the ellipses denote any extension to a full unitary on the orthogonal complements.
The explicit choice of the extensions does not affect the $2\times 2$ block in the relevant subspace.

A direct expansion gives
\beq
\begin{aligned}
U\hat W
&=
a_r e^{-\gamma_1 t}\ketb{r_1}{0^m}
+
e^{-\gamma_d t}\ketb{r_d}{0^{m-1}1}\\
&\quad-
a_r b_r e^{-\gamma_d t} e^{i(h_1-h_d)t}\ketb{r_d}{0^m}
+\cdots,
\end{aligned}
\eeq
and hence
\beq
\label{eq:VUW-appendix}
\hat V^\dagger U \hat W
=
a_r e^{-\gamma_1 t}\ketbra{0^m}
+
a_r e^{-\gamma_d t}G
+\cdots,
\eeq
where $G$ is supported on $\mathrm{span}\{\ket{0^m},\ket{0^{m-1}1}\}$ and, in the ordered basis
$\{\ket{0^m},\ket{0^{m-1}1}\}$, is
\beq
G=
\begin{pmatrix}
-|b_r|^2 e^{i(h_1-h_d)t} & b_r^*/a_r \\
-(b_r/a_r) e^{i(h_1-h_d)t} & 1-|b_r|^2
\end{pmatrix}.
\eeq
Finally,
\beq
\begin{aligned}
\|G\|_F^2 &= \sum_{i,j}|G_{ij}|^2\\
&=
|b_r|^4 + 2|b_r|^2(1-|b_r|^2) + (1-|b_r|^2)^2
&=1,
\end{aligned}
\eeq
so $\|G\|\le \|G\|_F=1$, independent of the angle between $\ket{r_1}$ and $\ket{r_d}$.
This is the bound used in \cref{sec:NHBQP} to control the postselection error in the complex-spectrum case.

\section{Single-Qubit $PT$-Symmetric $H$}
\label{app:arbitrary}

\subsection{$PT$-Symmetric $H$}

Here we show that any single-qubit quasi-Hermitian Hamiltonian (i.e., one for which there exists a positive definite metric $\eta>0$ satisfying $H^\dagger \eta=\eta H$, corresponding to the unbroken $PT$-symmetric phase)
is unitarily equivalent, up to an additive scalar and a choice of orthonormal basis, to the
canonical form used in \cref{sec:ptsym}.

First, write the traceless part of $H$ as
\beq
H \equiv H_0-i\Gamma,
\eeq
where $H_0$ and $\Gamma$ are Hermitian. Parameterize them as
\beq
H_0 \equiv g\vec n\cdot\vec\sigma,
\quad
\Gamma \equiv \gamma\vec m\cdot\vec\sigma,
\eeq
where $g,\gamma>0$, $\vec n,\vec m$ are unit vectors, and $\vec\sigma$ is the Pauli vector.
We can assume $\vec n$ and $\vec m$ are not parallel.
If $\vec m=\pm\vec n$, then $H=(g\mp i\gamma)\,\vec n\cdot\vec\sigma$ has eigenvalues $\pm(g\mp i\gamma)$, which are non-real for $\gamma>0$. Hence such $H$ is not quasi-Hermitian.

As recalled above, quasi-Hermiticity means that there exists $\eta>0$ such that
$H^\dag\eta=\eta H$.
Write $\eta$ in Bloch form as
\beq
\eta \equiv r_0\ident+\vec r\cdot\vec\sigma,
\eeq
with $r_0\in\mathbb{R}$, and $\vec r\in\mathbb{R}^3$ since $\eta$ is Hermitian.
Using $H^\dag=H_0+i\Gamma$, the condition becomes
\beq
0=H^\dag\eta-\eta H=[H_0,\eta]+i\{\Gamma,\eta\}.
\eeq
Using the Pauli identities
\beq
[\vec a\cdot\vec\sigma,\vec b\cdot\vec\sigma]=2i(\vec a\times\vec b)\cdot\vec\sigma,
\quad
\{\vec a\cdot\vec\sigma,\vec b\cdot\vec\sigma\}=2(\vec a\cdot\vec b)\ident,
\eeq
and the fact that $[\vec a\cdot\vec\sigma,\ident]=0$ and $\{\vec a\cdot\vec\sigma,\ident\}=2\vec a\cdot\vec\sigma$,
we obtain
\beq
\label{eq:C7}
0=g(\vec n\times\vec r)\cdot\vec\sigma+\gamma r_0\vec m\cdot\vec\sigma+\gamma(\vec m\cdot\vec r)\ident.
\eeq

Now choose an orthonormal basis $\{\vec v_1,\vec v_2,\vec v_3\}$ of $\mathbb{R}^3$ such that
$\vec n=\vec v_1$ and $\vec m=m_1\vec v_1+m_2\vec v_2$ (this is always possible since $\vec m$ is not parallel to $\vec n$).
Let $\sigma_i\equiv \vec v_i\cdot\vec\sigma$ and write $\vec r=r_1\vec v_1+r_2\vec v_2+r_3\vec v_3$, making $\vec n\times\vec r=r_2\vec v_3-r_3\vec v_2$. Equating coefficients of the Pauli basis in \cref{eq:C7} gives three equations from the Pauli components and one from the identity component:
\begin{subequations}
\label{eq:eta-constraints}
\begin{align}
\gamma r_0 m_1 &= 0 \\
-g r_3+\gamma r_0 m_2 &= 0 \\
g r_2 &= 0 \\
\vec m\cdot\vec r=m_1 r_1+m_2 r_2 &= 0.
\end{align}
\end{subequations}
Since $g,\gamma>0$, we have $r_2=0$. If $m_1\neq 0$ then the first, second, and fourth equations respectively force $r_0=r_3=r_1=0$, giving $\eta=0$, a contradiction. Therefore $m_1=0$,
so $\vec m=\vec v_2$ and $m_2=1$. In this basis,
\beq
H=g\sigma_1-i\gamma\sigma_2,
\quad
\eta=r_0\Bigl(\ident+\frac{\gamma}{g}\sigma_3\Bigr)+r_1\sigma_1.
\eeq
The eigenvalues of $\eta$ are $r_0\pm\sqrt{r_1^2+(r_0\gamma/g)^2}$, so $\eta>0$ implies $r_0>0$ and
\beq
\sqrt{\frac{\gamma^2}{g^2}+\frac{r_1^2}{r_0^2}}<1,
\eeq
which is only satisfiable when $\gamma<g$.

Finally, writing $H$ in the $\sigma_3$ eigenbasis yields \cref{eq:H-pt-sym}, the canonical single-qubit $PT$-symmetric form used in \cref{sec:ptsym}.

\subsection{Ellipsoidal Periodic $U$}

Since $\sigma_1$ and $\sigma_2$ anticommute, we have
$H^2=(g^2-\gamma^2)\ident$.
Define
$\omega \equiv \sqrt{g^2-\gamma^2}$,
which is strictly positive in the unbroken $PT$-symmetry regime $\gamma<g$. Then
\beq
U=e^{-iHt}=\cos(\omega t)\ident-\frac{i}{\omega}\sin(\omega t)H.
\eeq
At $t=\pi/(2\omega)$ this becomes $U=-(i/\omega)H$.
Geometrically, this periodic behavior reflects the fact that $PT$-symmetric evolution preserves the
norm induced by the metric $\eta>0$, corresponding to rotations on the associated ellipsoid in state space.
At the exceptional point $\gamma=g$, $\omega\to 0$, signaling the breakdown of periodic evolution.
Additionally, when $\gamma>g$, one has $\omega=i\sqrt{\gamma^2-g^2}$; the same closed form continues analytically and yields hyperbolic functions and exponential amplification/decay as one expects in the $PT$-broken phase.

\section{Trajectories Derivation}
\label{app:trajectories-purified}

Here we show how the Hamiltonian in \cref{eq:H-meter} generates quantum jumps and Lindblad evolution,
in addition to the no-jump trajectories discussed in the main text.

Recall that $H$ acts on system $A$ and meter $B$ as
\beq
H=H_A+\frac{1}{\sqrt{\delta}}H_{AB},
\eeq
and the timestep unitary is $U\equiv e^{-iH\delta}$. Expanding to the required order gives
\beq
U=\ident-i\sqrt{\delta}H_{AB}-i\delta H_A-\frac{\delta}{2}H_{AB}^2+O(\delta^{3/2}).
\eeq

\subsection{Quantum jumps}

Initialize the meter in $\ket{j}$ and postselect on $\ket{k}$ with $k\neq j$.
Define the effective (unnormalized) Kraus operator
\beq
C_{k\leftarrow j}\equiv \bra{k}U\ket{j}_B.
\eeq
Because $\bra{k}\ident\ket{j}=\bra{k}H_A\ket{j}=0$ for $k\neq j$, the leading contribution comes from
$H_{AB}$:
\beq
C_{k\leftarrow j}=-i\sqrt{\delta}\bra{k}H_{AB}\ket{j}_B+O(\delta).
\eeq
Using $H_{AB}=\sum_{a\neq b}L_{ab}\otimes\ket{a}\bra{b}$, we have $\bra{k}H_{AB}\ket{j}_B=L_{kj}$, hence
\beq
C_{k\leftarrow j}=-i\sqrt{\delta}L_{kj}+O(\delta).
\eeq
Therefore the postselected (unnormalized) state after one step is
\beq
\tilde{\mathcal{E}}_{k\leftarrow j}(\rho)\equiv C_{k\leftarrow j}\rho C_{k\leftarrow j}^\dag
=\delta L_{kj}\rho L_{kj}^\dag+O(\delta^{3/2}),
\eeq
which is the jump form stated in \cref{eq:jump}. The scaling $H_{AB}/\sqrt{\delta}$ ensures that
the jump probability per step, $\Tr[\tilde{\mathcal{E}}_{k\leftarrow j}(\rho)]$, is $O(\delta)$, matching the usual quantum-trajectories limit.

\subsection{Lindblad evolution from tracing out the meter}

Now replace postselection by a partial trace over the meter:
\beq
\rho(\delta)\equiv \mathcal{E}(\rho)=\Tr_B\Bigl[U(\rho\otimes\ketb{j}{j})U^\dag\Bigr].
\eeq
Since this operation is trace preserving, $\tilde{\mathcal{E}} = \mathcal{E}$, and the normalization step can be dropped. Decompose the trace over the meter basis:
\beq
\mathcal{E}(\rho)=\bra{j}U(\rho\otimes\ketb{j}{j})U^\dag\ket{j}_B+\sum_{k\neq j}\bra{k}U(\rho\otimes\ketb{j}{j})U^\dag\ket{k}_B.
\eeq
The $k\neq j$ terms reproduce the jump contributions $\delta L_{kj}\rho L_{kj}^\dag$ to leading order
as above, while the $j\to j$ block produces the coherent drift and the anticommutator term from
$H_{AB}^2$, described in \cref{eq:H_0_eff,eq:H_k_eff}. Collecting terms yields
\beq
\begin{aligned}
\rho(\delta)
&=\rho-i\delta[H_j,\rho]
-\frac{\delta}{2}\sum_{k\neq j}\{L_{kj}^\dag L_{kj},\rho\}
+\delta\sum_{k\neq j}L_{kj}\rho L_{kj}^\dag\\
&\quad +O(\delta^{3/2}),
\end{aligned}
\eeq
which in differential form gives the Lindblad equation with system Hamiltonian $H_j$ in the limit
$\delta\to 0$.

\section{Relation to the $\mathrm{BNQP}$ model~\cite{zhang2025physics}}
\label{app:ZhangWu}

In related work that appeared after the first version of our preprint, Zhang \& Wu~\cite{zhang2025physics}
introduce a model of non-unitary quantum computing that augments a standard universal unitary gate set with a
fixed non-unitary single-qubit gate $G=\mathrm{diag}(g,g^{-1})$, for $g>0,\ g\neq 1$.
They define the corresponding bounded-error class $\mathrm{BNQP}$ and argue that
$\mathrm{BNQP}=\mathrm{P}^{\#\mathrm{P}}$.

At the level of normalized output statistics without adaptive classical access to a $\mathrm{PP}$ routine, their fixed-gate model coincides with the special
case of our framework $\mathrm{NHBQP}(U)$ with $U=G$: since renormalization only multiplies the state by an
overall scalar, it commutes through subsequent linear gates and cancels upon final normalization, and their
definition does not allow intermediate measurements followed by further gates. In our uniform circuit-family
setting (polynomial-size circuits with renormalization and no external oracle access), and using \cref{thm:NHBQP-characterization}, we thus obtain $\mathrm{BNQP}=\mathrm{NHBQP}(G)=\mathrm{PostBQP}=\mathrm{PP}$.
Accordingly, oracle-class expressions such as $\mathrm{P}^{\#\mathrm{P}}=\mathrm{P}^{\mathrm{PP}}$ must involve an
additional assumption of classical polynomial-time computation with adaptive oracle access to a
$\mathrm{PP}$ (equivalently $\mathrm{PostBQP}$) subroutine, which we do not include in the definition of
$\mathrm{NHBQP}(U)$.

To pinpoint where the stronger $\mathrm{BNQP}=\mathrm{P}^{\#\mathrm{P}}$-type conclusion can enter, note that the route in
Ref.~\cite{zhang2025physics} is explicitly motivated by their earlier Lorentz-quantum-computer (LQC)
construction~\cite{zhang2026power}, where they first propose a bounded-error procedure for a $\mathrm{PP}$-complete
language (MAJSAT), and then invoke the standard identity $\mathrm{P}^{\mathrm{PP}}=\mathrm{P}^{\#\mathrm{P}}$.

Specifically, in Ref.~\cite{zhang2026power} the authors discuss a problem they call \textsc{MAX-$k$-Independent-Set}
as an example of a $\mathrm{P}^{\#\mathrm{P}}$ (equivalently $\mathrm{P}^{\mathrm{PP}}$) task, and give a procedure that uses
polynomially many oracular calls to their $\mathrm{PP}$ subroutine for MAJSAT.
This is consistent with $\mathrm{P}^{\mathrm{PP}}$, but it relies on an additional oracle assumption
(adaptive classical access to a $\mathrm{PP}$ subroutine) that is not part of the standard uniform circuit-family definition
of $\mathrm{BNQP}$ (nor of our $\mathrm{NHBQP}(U)$).
Note that while \textsc{MAX-$k$-IS} $\in\mathrm{P}^{\#\mathrm{P}}$, we are not aware of a proof that the particular \textsc{MAX-$k$-IS} formulation used there is $\mathrm{P}^{\#\mathrm{P}}$-complete or $\mathrm{P}^{\#\mathrm{P}}$-hard.

Since the subsequent $\mathrm{P}^{\#\mathrm{P}}$ discussion in~\cite{zhang2025physics} is built on this $\mathrm{PP}$
subroutine and on oracle-class reasoning, we view $\mathrm{P}^{\#\mathrm{P}}$ as reflecting either (i) an
oracle-augmented model (adaptive classical access to a $\mathrm{PP}$ routine, absent in our work) or (ii) an implicit promise/extra
assumption not present in the standard uniform circuit-family definition of $\mathrm{BNQP}$.

In additional contrast, our results treat arbitrary non-unitary gates (and size-dependent families $\{U_n\}$ under
explicit uniformity/implementability assumptions), and we further develop a purification-based simulability
framework for identifying efficiently simulable NH dynamics. Zhang \& Wu~\cite{zhang2025physics} also analyze
physical schemes for implementing $G$ and argue that the required physical resources scale exponentially,
supporting the view that the apparent complexity-theoretic power of fixed non-unitary gates is unlikely to be
scalable in a universal quantum-computing setting.
\end{appendix}

\bibliographystyle{apsrev4-1}
\bibliography{bib}

\begin{thebibliography}{77}%
\makeatletter
\providecommand \@ifxundefined [1]{%
 \@ifx{#1\undefined}
}%
\providecommand \@ifnum [1]{%
 \ifnum #1\expandafter \@firstoftwo
 \else \expandafter \@secondoftwo
 \fi
}%
\providecommand \@ifx [1]{%
 \ifx #1\expandafter \@firstoftwo
 \else \expandafter \@secondoftwo
 \fi
}%
\providecommand \natexlab [1]{#1}%
\providecommand \enquote  [1]{``#1''}%
\providecommand \bibnamefont  [1]{#1}%
\providecommand \bibfnamefont [1]{#1}%
\providecommand \citenamefont [1]{#1}%
\providecommand \href@noop [0]{\@secondoftwo}%
\providecommand \href [0]{\begingroup \@sanitize@url \@href}%
\providecommand \@href[1]{\@@startlink{#1}\@@href}%
\providecommand \@@href[1]{\endgroup#1\@@endlink}%
\providecommand \@sanitize@url [0]{\catcode `\\12\catcode `\$12\catcode
  `\&12\catcode `\#12\catcode `\^12\catcode `\_12\catcode `\%12\relax}%
\providecommand \@@startlink[1]{}%
\providecommand \@@endlink[0]{}%
\providecommand \url  [0]{\begingroup\@sanitize@url \@url }%
\providecommand \@url [1]{\endgroup\@href {#1}{\urlprefix }}%
\providecommand \urlprefix  [0]{URL }%
\providecommand \Eprint [0]{\href }%
\providecommand \doibase [0]{http://dx.doi.org/}%
\providecommand \selectlanguage [0]{\@gobble}%
\providecommand \bibinfo  [0]{\@secondoftwo}%
\providecommand \bibfield  [0]{\@secondoftwo}%
\providecommand \translation [1]{[#1]}%
\providecommand \BibitemOpen [0]{}%
\providecommand \bibitemStop [0]{}%
\providecommand \bibitemNoStop [0]{.\EOS\space}%
\providecommand \EOS [0]{\spacefactor3000\relax}%
\providecommand \BibitemShut  [1]{\csname bibitem#1\endcsname}%
\let\auto@bib@innerbib\@empty
\bibitem [{\citenamefont {Ashida}\ \emph {et~al.}(2020)\citenamefont {Ashida},
  \citenamefont {Gong},\ and\ \citenamefont {Ueda}}]{ashida2020nonhermitian}%
  \BibitemOpen
  \bibfield  {author} {\bibinfo {author} {\bibfnamefont {Y.}~\bibnamefont
  {Ashida}}, \bibinfo {author} {\bibfnamefont {Z.}~\bibnamefont {Gong}}, \ and\
  \bibinfo {author} {\bibfnamefont {M.}~\bibnamefont {Ueda}},\ }\href {\doibase
  10.1080/00018732.2021.1876991} {\bibfield  {journal} {\bibinfo  {journal}
  {Advances in Physics}\ }\textbf {\bibinfo {volume} {69}},\ \bibinfo {pages}
  {249} (\bibinfo {year} {2020})}\BibitemShut {NoStop}%
\bibitem [{\citenamefont {Ashida}\ and\ \citenamefont
  {Ueda}(2018)}]{ashida2018full}%
  \BibitemOpen
  \bibfield  {author} {\bibinfo {author} {\bibfnamefont {Y.}~\bibnamefont
  {Ashida}}\ and\ \bibinfo {author} {\bibfnamefont {M.}~\bibnamefont {Ueda}},\
  }\href {\doibase 10.1103/physrevlett.120.185301} {\bibfield  {journal}
  {\bibinfo  {journal} {Physical Review Letters}\ }\textbf {\bibinfo {volume}
  {120}},\ \bibinfo {pages} {185301} (\bibinfo {year} {2018})}\BibitemShut
  {NoStop}%
\bibitem [{\citenamefont {Bender}\ and\ \citenamefont
  {Boettcher}(1998)}]{Bender_1998_real_spectra}%
  \BibitemOpen
  \bibfield  {author} {\bibinfo {author} {\bibfnamefont {C.~M.}\ \bibnamefont
  {Bender}}\ and\ \bibinfo {author} {\bibfnamefont {S.}~\bibnamefont
  {Boettcher}},\ }\href {\doibase 10.1103/PhysRevLett.80.5243} {\bibfield
  {journal} {\bibinfo  {journal} {Phys. Rev. Lett.}\ }\textbf {\bibinfo
  {volume} {80}},\ \bibinfo {pages} {5243} (\bibinfo {year}
  {1998})}\BibitemShut {NoStop}%
\bibitem [{\citenamefont {Bender}\ \emph {et~al.}(2002)\citenamefont {Bender},
  \citenamefont {Brody},\ and\ \citenamefont
  {Jones}}]{Bender_2002_complex_extension}%
  \BibitemOpen
  \bibfield  {author} {\bibinfo {author} {\bibfnamefont {C.~M.}\ \bibnamefont
  {Bender}}, \bibinfo {author} {\bibfnamefont {D.~C.}\ \bibnamefont {Brody}}, \
  and\ \bibinfo {author} {\bibfnamefont {H.~F.}\ \bibnamefont {Jones}},\ }\href
  {\doibase 10.1103/PhysRevLett.89.270401} {\bibfield  {journal} {\bibinfo
  {journal} {Phys. Rev. Lett.}\ }\textbf {\bibinfo {volume} {89}},\ \bibinfo
  {pages} {270401} (\bibinfo {year} {2002})}\BibitemShut {NoStop}%
\bibitem [{\citenamefont
  {Mostafazadeh}(2002)}]{mostafazadeh_pseudo-hermiticity_2002}%
  \BibitemOpen
  \bibfield  {author} {\bibinfo {author} {\bibfnamefont {A.}~\bibnamefont
  {Mostafazadeh}},\ }\href {\doibase 10.1063/1.1418246} {\bibfield  {journal}
  {\bibinfo  {journal} {Journal of Mathematical Physics}\ }\textbf {\bibinfo
  {volume} {43}},\ \bibinfo {pages} {205} (\bibinfo {year} {2002})}\BibitemShut
  {NoStop}%
\bibitem [{\citenamefont {Mostafazadeh}(2003)}]{mostafazadeh_2003}%
  \BibitemOpen
  \bibfield  {author} {\bibinfo {author} {\bibfnamefont {A.}~\bibnamefont
  {Mostafazadeh}},\ }\href {\doibase 10.1023/b:cjop.0000010537.23790.8c}
  {\bibfield  {journal} {\bibinfo  {journal} {Czechoslovak Journal of Physics}\
  }\textbf {\bibinfo {volume} {53}},\ \bibinfo {pages} {1079} (\bibinfo {year}
  {2003})}\BibitemShut {NoStop}%
\bibitem [{\citenamefont {Wu}\ \emph {et~al.}(2019)\citenamefont {Wu},
  \citenamefont {Liu}, \citenamefont {Geng}, \citenamefont {Song},
  \citenamefont {Ye}, \citenamefont {Yu}, \citenamefont {Peng},\ and\
  \citenamefont {Du}}]{wu2019observation}%
  \BibitemOpen
  \bibfield  {author} {\bibinfo {author} {\bibfnamefont {Y.}~\bibnamefont
  {Wu}}, \bibinfo {author} {\bibfnamefont {W.}~\bibnamefont {Liu}}, \bibinfo
  {author} {\bibfnamefont {J.}~\bibnamefont {Geng}}, \bibinfo {author}
  {\bibfnamefont {X.}~\bibnamefont {Song}}, \bibinfo {author} {\bibfnamefont
  {X.}~\bibnamefont {Ye}}, \bibinfo {author} {\bibfnamefont {C.}~\bibnamefont
  {Yu}}, \bibinfo {author} {\bibfnamefont {X.}~\bibnamefont {Peng}}, \ and\
  \bibinfo {author} {\bibfnamefont {J.}~\bibnamefont {Du}},\ }\href {\doibase
  10.1126/science.aaw8205} {\bibfield  {journal} {\bibinfo  {journal}
  {Science}\ }\textbf {\bibinfo {volume} {364}},\ \bibinfo {pages} {878}
  (\bibinfo {year} {2019})}\BibitemShut {NoStop}%
\bibitem [{\citenamefont {Li}\ \emph {et~al.}(2019{\natexlab{a}})\citenamefont
  {Li}, \citenamefont {Harter}, \citenamefont {Liu}, \citenamefont {de~Melo},
  \citenamefont {Joglekar},\ and\ \citenamefont {Luo}}]{li2019observation}%
  \BibitemOpen
  \bibfield  {author} {\bibinfo {author} {\bibfnamefont {J.}~\bibnamefont
  {Li}}, \bibinfo {author} {\bibfnamefont {A.~K.}\ \bibnamefont {Harter}},
  \bibinfo {author} {\bibfnamefont {J.}~\bibnamefont {Liu}}, \bibinfo {author}
  {\bibfnamefont {L.}~\bibnamefont {de~Melo}}, \bibinfo {author} {\bibfnamefont
  {Y.~N.}\ \bibnamefont {Joglekar}}, \ and\ \bibinfo {author} {\bibfnamefont
  {L.}~\bibnamefont {Luo}},\ }\href {\doibase 10.1038/s41467-019-08596-1}
  {\bibfield  {journal} {\bibinfo  {journal} {Nature Communications}\ }\textbf
  {\bibinfo {volume} {10}},\ \bibinfo {pages} {855} (\bibinfo {year}
  {2019}{\natexlab{a}})}\BibitemShut {NoStop}%
\bibitem [{\citenamefont {Naghiloo}\ \emph {et~al.}(2019)\citenamefont
  {Naghiloo}, \citenamefont {Abbasi}, \citenamefont {Joglekar},\ and\
  \citenamefont {Murch}}]{naghiloo2019EP}%
  \BibitemOpen
  \bibfield  {author} {\bibinfo {author} {\bibfnamefont {M.}~\bibnamefont
  {Naghiloo}}, \bibinfo {author} {\bibfnamefont {M.}~\bibnamefont {Abbasi}},
  \bibinfo {author} {\bibfnamefont {Y.~N.}\ \bibnamefont {Joglekar}}, \ and\
  \bibinfo {author} {\bibfnamefont {K.~W.}\ \bibnamefont {Murch}},\ }\href
  {\doibase 10.1038/s41567-019-0652-z} {\bibfield  {journal} {\bibinfo
  {journal} {Nature Physics}\ }\textbf {\bibinfo {volume} {15}},\ \bibinfo
  {pages} {1232} (\bibinfo {year} {2019})}\BibitemShut {NoStop}%
\bibitem [{\citenamefont {R{\"u}ter}\ \emph {et~al.}(2010)\citenamefont
  {R{\"u}ter}, \citenamefont {Makris}, \citenamefont {El-Ganainy},
  \citenamefont {Christodoulides}, \citenamefont {Segev},\ and\ \citenamefont
  {Kip}}]{ruter2010PT}%
  \BibitemOpen
  \bibfield  {author} {\bibinfo {author} {\bibfnamefont {C.~E.}\ \bibnamefont
  {R{\"u}ter}}, \bibinfo {author} {\bibfnamefont {K.~G.}\ \bibnamefont
  {Makris}}, \bibinfo {author} {\bibfnamefont {R.}~\bibnamefont {El-Ganainy}},
  \bibinfo {author} {\bibfnamefont {D.~N.}\ \bibnamefont {Christodoulides}},
  \bibinfo {author} {\bibfnamefont {M.}~\bibnamefont {Segev}}, \ and\ \bibinfo
  {author} {\bibfnamefont {D.}~\bibnamefont {Kip}},\ }\href {\doibase
  10.1038/nphys1515} {\bibfield  {journal} {\bibinfo  {journal} {Nature
  Physics}\ }\textbf {\bibinfo {volume} {6}},\ \bibinfo {pages} {192} (\bibinfo
  {year} {2010})}\BibitemShut {NoStop}%
\bibitem [{\citenamefont {Regensburger}\ \emph {et~al.}(2012)\citenamefont
  {Regensburger}, \citenamefont {Bersch}, \citenamefont {Miri}, \citenamefont
  {Onishchukov}, \citenamefont {Christodoulides},\ and\ \citenamefont
  {Peschel}}]{regensburger2012PT}%
  \BibitemOpen
  \bibfield  {author} {\bibinfo {author} {\bibfnamefont {A.}~\bibnamefont
  {Regensburger}}, \bibinfo {author} {\bibfnamefont {C.}~\bibnamefont
  {Bersch}}, \bibinfo {author} {\bibfnamefont {M.-A.}\ \bibnamefont {Miri}},
  \bibinfo {author} {\bibfnamefont {G.}~\bibnamefont {Onishchukov}}, \bibinfo
  {author} {\bibfnamefont {D.~N.}\ \bibnamefont {Christodoulides}}, \ and\
  \bibinfo {author} {\bibfnamefont {U.}~\bibnamefont {Peschel}},\ }\href
  {\doibase 10.1038/nature11298} {\bibfield  {journal} {\bibinfo  {journal}
  {Nature}\ }\textbf {\bibinfo {volume} {488}},\ \bibinfo {pages} {167}
  (\bibinfo {year} {2012})}\BibitemShut {NoStop}%
\bibitem [{\citenamefont {Zeuner}\ \emph {et~al.}(2015)\citenamefont {Zeuner},
  \citenamefont {Rechtsman}, \citenamefont {Plotnik}, \citenamefont {Lumer},
  \citenamefont {Nolte}, \citenamefont {Rudner}, \citenamefont {Segev},\ and\
  \citenamefont {Szameit}}]{zeuner2015topological}%
  \BibitemOpen
  \bibfield  {author} {\bibinfo {author} {\bibfnamefont {J.~M.}\ \bibnamefont
  {Zeuner}}, \bibinfo {author} {\bibfnamefont {M.~C.}\ \bibnamefont
  {Rechtsman}}, \bibinfo {author} {\bibfnamefont {Y.}~\bibnamefont {Plotnik}},
  \bibinfo {author} {\bibfnamefont {Y.}~\bibnamefont {Lumer}}, \bibinfo
  {author} {\bibfnamefont {S.}~\bibnamefont {Nolte}}, \bibinfo {author}
  {\bibfnamefont {M.~S.}\ \bibnamefont {Rudner}}, \bibinfo {author}
  {\bibfnamefont {M.}~\bibnamefont {Segev}}, \ and\ \bibinfo {author}
  {\bibfnamefont {A.}~\bibnamefont {Szameit}},\ }\href {\doibase
  10.1103/PhysRevLett.115.040402} {\bibfield  {journal} {\bibinfo  {journal}
  {Physical Review Letters}\ }\textbf {\bibinfo {volume} {115}},\ \bibinfo
  {pages} {040402} (\bibinfo {year} {2015})}\BibitemShut {NoStop}%
\bibitem [{\citenamefont {Hodaei}\ \emph {et~al.}(2015)\citenamefont {Hodaei},
  \citenamefont {Miri}, \citenamefont {Hassan}, \citenamefont {Hayenga},
  \citenamefont {Heinrich}, \citenamefont {Christodoulides},\ and\
  \citenamefont {Khajavikhan}}]{hodaei2015tunable}%
  \BibitemOpen
  \bibfield  {author} {\bibinfo {author} {\bibfnamefont {H.}~\bibnamefont
  {Hodaei}}, \bibinfo {author} {\bibfnamefont {M.-A.}\ \bibnamefont {Miri}},
  \bibinfo {author} {\bibfnamefont {A.~U.}\ \bibnamefont {Hassan}}, \bibinfo
  {author} {\bibfnamefont {W.}~\bibnamefont {Hayenga}}, \bibinfo {author}
  {\bibfnamefont {M.}~\bibnamefont {Heinrich}}, \bibinfo {author}
  {\bibfnamefont {D.~N.}\ \bibnamefont {Christodoulides}}, \ and\ \bibinfo
  {author} {\bibfnamefont {M.}~\bibnamefont {Khajavikhan}},\ }\href {\doibase
  10.1364/OL.40.004955} {\bibfield  {journal} {\bibinfo  {journal} {Optics
  Letters}\ }\textbf {\bibinfo {volume} {40}},\ \bibinfo {pages} {4955}
  (\bibinfo {year} {2015})}\BibitemShut {NoStop}%
\bibitem [{\citenamefont {Gopalakrishnan}\ and\ \citenamefont
  {Gullans}(2021)}]{gopalakrishnan_entanglement_2021}%
  \BibitemOpen
  \bibfield  {author} {\bibinfo {author} {\bibfnamefont {S.}~\bibnamefont
  {Gopalakrishnan}}\ and\ \bibinfo {author} {\bibfnamefont {M.~J.}\
  \bibnamefont {Gullans}},\ }\href {\doibase 10.1103/PhysRevLett.126.170503}
  {\bibfield  {journal} {\bibinfo  {journal} {Physical Review Letters}\
  }\textbf {\bibinfo {volume} {126}},\ \bibinfo {pages} {170503} (\bibinfo
  {year} {2021})}\BibitemShut {NoStop}%
\bibitem [{\citenamefont {Matsumoto}\ \emph {et~al.}(2020)\citenamefont
  {Matsumoto}, \citenamefont {Kawabata}, \citenamefont {Ashida}, \citenamefont
  {Furukawa},\ and\ \citenamefont {Ueda}}]{matsumoto2020continuous}%
  \BibitemOpen
  \bibfield  {author} {\bibinfo {author} {\bibfnamefont {N.}~\bibnamefont
  {Matsumoto}}, \bibinfo {author} {\bibfnamefont {K.}~\bibnamefont {Kawabata}},
  \bibinfo {author} {\bibfnamefont {Y.}~\bibnamefont {Ashida}}, \bibinfo
  {author} {\bibfnamefont {S.}~\bibnamefont {Furukawa}}, \ and\ \bibinfo
  {author} {\bibfnamefont {M.}~\bibnamefont {Ueda}},\ }\href {\doibase
  10.1103/PhysRevLett.125.260601} {\bibfield  {journal} {\bibinfo  {journal}
  {Phys. Rev. Lett.}\ }\textbf {\bibinfo {volume} {125}},\ \bibinfo {pages}
  {260601} (\bibinfo {year} {2020})}\BibitemShut {NoStop}%
\bibitem [{\citenamefont {Li}\ \emph {et~al.}(2018)\citenamefont {Li},
  \citenamefont {Chen},\ and\ \citenamefont {Fisher}}]{PhysRevB.98.205136}%
  \BibitemOpen
  \bibfield  {author} {\bibinfo {author} {\bibfnamefont {Y.}~\bibnamefont
  {Li}}, \bibinfo {author} {\bibfnamefont {X.}~\bibnamefont {Chen}}, \ and\
  \bibinfo {author} {\bibfnamefont {M.~P.~A.}\ \bibnamefont {Fisher}},\ }\href
  {\doibase 10.1103/PhysRevB.98.205136} {\bibfield  {journal} {\bibinfo
  {journal} {Phys. Rev. B}\ }\textbf {\bibinfo {volume} {98}},\ \bibinfo
  {pages} {205136} (\bibinfo {year} {2018})}\BibitemShut {NoStop}%
\bibitem [{\citenamefont {Skinner}\ \emph {et~al.}(2019)\citenamefont
  {Skinner}, \citenamefont {Ruhman},\ and\ \citenamefont
  {Nahum}}]{PhysRevX.9.031009}%
  \BibitemOpen
  \bibfield  {author} {\bibinfo {author} {\bibfnamefont {B.}~\bibnamefont
  {Skinner}}, \bibinfo {author} {\bibfnamefont {J.}~\bibnamefont {Ruhman}}, \
  and\ \bibinfo {author} {\bibfnamefont {A.}~\bibnamefont {Nahum}},\ }\href
  {\doibase 10.1103/PhysRevX.9.031009} {\bibfield  {journal} {\bibinfo
  {journal} {Phys. Rev. X}\ }\textbf {\bibinfo {volume} {9}},\ \bibinfo {pages}
  {031009} (\bibinfo {year} {2019})}\BibitemShut {NoStop}%
\bibitem [{\citenamefont {Chan}\ \emph {et~al.}(2019)\citenamefont {Chan},
  \citenamefont {Nandkishore}, \citenamefont {Pretko},\ and\ \citenamefont
  {Smith}}]{PhysRevB.99.224307}%
  \BibitemOpen
  \bibfield  {author} {\bibinfo {author} {\bibfnamefont {A.}~\bibnamefont
  {Chan}}, \bibinfo {author} {\bibfnamefont {R.~M.}\ \bibnamefont
  {Nandkishore}}, \bibinfo {author} {\bibfnamefont {M.}~\bibnamefont {Pretko}},
  \ and\ \bibinfo {author} {\bibfnamefont {G.}~\bibnamefont {Smith}},\ }\href
  {\doibase 10.1103/PhysRevB.99.224307} {\bibfield  {journal} {\bibinfo
  {journal} {Phys. Rev. B}\ }\textbf {\bibinfo {volume} {99}},\ \bibinfo
  {pages} {224307} (\bibinfo {year} {2019})}\BibitemShut {NoStop}%
\bibitem [{\citenamefont {Li}\ \emph {et~al.}(2019{\natexlab{b}})\citenamefont
  {Li}, \citenamefont {Chen},\ and\ \citenamefont
  {Fisher}}]{PhysRevB.100.134306}%
  \BibitemOpen
  \bibfield  {author} {\bibinfo {author} {\bibfnamefont {Y.}~\bibnamefont
  {Li}}, \bibinfo {author} {\bibfnamefont {X.}~\bibnamefont {Chen}}, \ and\
  \bibinfo {author} {\bibfnamefont {M.~P.~A.}\ \bibnamefont {Fisher}},\ }\href
  {\doibase 10.1103/PhysRevB.100.134306} {\bibfield  {journal} {\bibinfo
  {journal} {Phys. Rev. B}\ }\textbf {\bibinfo {volume} {100}},\ \bibinfo
  {pages} {134306} (\bibinfo {year} {2019}{\natexlab{b}})}\BibitemShut
  {NoStop}%
\bibitem [{\citenamefont {Gullans}\ and\ \citenamefont
  {Huse}(2020)}]{PhysRevX.10.041020}%
  \BibitemOpen
  \bibfield  {author} {\bibinfo {author} {\bibfnamefont {M.~J.}\ \bibnamefont
  {Gullans}}\ and\ \bibinfo {author} {\bibfnamefont {D.~A.}\ \bibnamefont
  {Huse}},\ }\href {\doibase 10.1103/PhysRevX.10.041020} {\bibfield  {journal}
  {\bibinfo  {journal} {Phys. Rev. X}\ }\textbf {\bibinfo {volume} {10}},\
  \bibinfo {pages} {041020} (\bibinfo {year} {2020})}\BibitemShut {NoStop}%
\bibitem [{\citenamefont {Choi}\ \emph {et~al.}(2020)\citenamefont {Choi},
  \citenamefont {Bao}, \citenamefont {Qi},\ and\ \citenamefont
  {Altman}}]{PhysRevLett.125.030505}%
  \BibitemOpen
  \bibfield  {author} {\bibinfo {author} {\bibfnamefont {S.}~\bibnamefont
  {Choi}}, \bibinfo {author} {\bibfnamefont {Y.}~\bibnamefont {Bao}}, \bibinfo
  {author} {\bibfnamefont {X.-L.}\ \bibnamefont {Qi}}, \ and\ \bibinfo {author}
  {\bibfnamefont {E.}~\bibnamefont {Altman}},\ }\href {\doibase
  10.1103/PhysRevLett.125.030505} {\bibfield  {journal} {\bibinfo  {journal}
  {Phys. Rev. Lett.}\ }\textbf {\bibinfo {volume} {125}},\ \bibinfo {pages}
  {030505} (\bibinfo {year} {2020})}\BibitemShut {NoStop}%
\bibitem [{\citenamefont {Ippoliti}\ \emph {et~al.}(2021)\citenamefont
  {Ippoliti}, \citenamefont {Gullans}, \citenamefont {Gopalakrishnan},
  \citenamefont {Huse},\ and\ \citenamefont {Khemani}}]{PhysRevX.11.011030}%
  \BibitemOpen
  \bibfield  {author} {\bibinfo {author} {\bibfnamefont {M.}~\bibnamefont
  {Ippoliti}}, \bibinfo {author} {\bibfnamefont {M.~J.}\ \bibnamefont
  {Gullans}}, \bibinfo {author} {\bibfnamefont {S.}~\bibnamefont
  {Gopalakrishnan}}, \bibinfo {author} {\bibfnamefont {D.~A.}\ \bibnamefont
  {Huse}}, \ and\ \bibinfo {author} {\bibfnamefont {V.}~\bibnamefont
  {Khemani}},\ }\href {\doibase 10.1103/PhysRevX.11.011030} {\bibfield
  {journal} {\bibinfo  {journal} {Phys. Rev. X}\ }\textbf {\bibinfo {volume}
  {11}},\ \bibinfo {pages} {011030} (\bibinfo {year} {2021})}\BibitemShut
  {NoStop}%
\bibitem [{\citenamefont {Das~Agarwal}\ \emph {et~al.}(2026)\citenamefont
  {Das~Agarwal}, \citenamefont {Konar}, \citenamefont {Lakkaraju},\ and\
  \citenamefont {Sen(De)}}]{agarwal2026recognizing}%
  \BibitemOpen
  \bibfield  {author} {\bibinfo {author} {\bibfnamefont {K.}~\bibnamefont
  {Das~Agarwal}}, \bibinfo {author} {\bibfnamefont {T.~K.}\ \bibnamefont
  {Konar}}, \bibinfo {author} {\bibfnamefont {L.~G.~C.}\ \bibnamefont
  {Lakkaraju}}, \ and\ \bibinfo {author} {\bibfnamefont {A.}~\bibnamefont
  {Sen(De)}},\ }\href {\doibase 10.1103/b5sd-fn57} {\bibfield  {journal}
  {\bibinfo  {journal} {Physical Review A}\ }\textbf {\bibinfo {volume} {113}}
  (\bibinfo {year} {2026}),\ 10.1103/b5sd-fn57}\BibitemShut {NoStop}%
\bibitem [{\citenamefont {Barch}\ \emph {et~al.}(2023)\citenamefont {Barch},
  \citenamefont {Anand}, \citenamefont {Marshall}, \citenamefont {Rieffel},\
  and\ \citenamefont {Zanardi}}]{barch2023scrambling}%
  \BibitemOpen
  \bibfield  {author} {\bibinfo {author} {\bibfnamefont {B.}~\bibnamefont
  {Barch}}, \bibinfo {author} {\bibfnamefont {N.}~\bibnamefont {Anand}},
  \bibinfo {author} {\bibfnamefont {J.}~\bibnamefont {Marshall}}, \bibinfo
  {author} {\bibfnamefont {E.}~\bibnamefont {Rieffel}}, \ and\ \bibinfo
  {author} {\bibfnamefont {P.}~\bibnamefont {Zanardi}},\ }\href {\doibase
  10.1103/PhysRevB.108.134305} {\bibfield  {journal} {\bibinfo  {journal}
  {Phys. Rev. B}\ }\textbf {\bibinfo {volume} {108}},\ \bibinfo {pages}
  {134305} (\bibinfo {year} {2023})}\BibitemShut {NoStop}%
\bibitem [{\citenamefont {Barch}(2024)}]{barch2024locality}%
  \BibitemOpen
  \bibfield  {author} {\bibinfo {author} {\bibfnamefont {B.}~\bibnamefont
  {Barch}},\ }\href {\doibase 10.1103/PhysRevB.110.094307} {\bibfield
  {journal} {\bibinfo  {journal} {Phys. Rev. B}\ }\textbf {\bibinfo {volume}
  {110}},\ \bibinfo {pages} {094307} (\bibinfo {year} {2024})}\BibitemShut
  {NoStop}%
\bibitem [{\citenamefont {Zhang}\ \emph {et~al.}(2025)\citenamefont {Zhang},
  \citenamefont {Carrasquilla},\ and\ \citenamefont
  {Kim}}]{zhang2025observation}%
  \BibitemOpen
  \bibfield  {author} {\bibinfo {author} {\bibfnamefont {Y.}~\bibnamefont
  {Zhang}}, \bibinfo {author} {\bibfnamefont {J.}~\bibnamefont {Carrasquilla}},
  \ and\ \bibinfo {author} {\bibfnamefont {Y.~B.}\ \bibnamefont {Kim}},\ }\href
  {\doibase 10.1038/s41467-025-57930-3} {\bibfield  {journal} {\bibinfo
  {journal} {Nature Communications}\ }\textbf {\bibinfo {volume} {16}}
  (\bibinfo {year} {2025}),\ 10.1038/s41467-025-57930-3}\BibitemShut {NoStop}%
\bibitem [{\citenamefont {Abrams}\ and\ \citenamefont
  {Lloyd}(1998)}]{abrams1998nonlinear}%
  \BibitemOpen
  \bibfield  {author} {\bibinfo {author} {\bibfnamefont {D.~S.}\ \bibnamefont
  {Abrams}}\ and\ \bibinfo {author} {\bibfnamefont {S.}~\bibnamefont {Lloyd}},\
  }\href {\doibase 10.1103/physrevlett.81.3992} {\bibfield  {journal} {\bibinfo
   {journal} {Physical Review Letters}\ }\textbf {\bibinfo {volume} {81}},\
  \bibinfo {pages} {3992} (\bibinfo {year} {1998})}\BibitemShut {NoStop}%
\bibitem [{\citenamefont {Mochizuki}\ and\ \citenamefont
  {Hamazaki}(2023)}]{Mochizuki2023}%
  \BibitemOpen
  \bibfield  {author} {\bibinfo {author} {\bibfnamefont {K.}~\bibnamefont
  {Mochizuki}}\ and\ \bibinfo {author} {\bibfnamefont {R.}~\bibnamefont
  {Hamazaki}},\ }\href {\doibase 10.1103/PhysRevResearch.5.013177} {\bibfield
  {journal} {\bibinfo  {journal} {Phys. Rev. Res.}\ }\textbf {\bibinfo {volume}
  {5}},\ \bibinfo {pages} {013177} (\bibinfo {year} {2023})}\BibitemShut
  {NoStop}%
\bibitem [{\citenamefont {Aaronson}(2005)}]{Aaronson:2005aa}%
  \BibitemOpen
  \bibfield  {author} {\bibinfo {author} {\bibfnamefont {S.}~\bibnamefont
  {Aaronson}},\ }\href {\doibase 10.1098/rspa.2005.1546} {\bibfield  {journal}
  {\bibinfo  {journal} {Proceedings of the Royal Society A: Mathematical,
  Physical and Engineering Sciences}\ }\textbf {\bibinfo {volume} {461}},\
  \bibinfo {pages} {3473} (\bibinfo {year} {2005})}\BibitemShut {NoStop}%
\bibitem [{\citenamefont {Bender}\ \emph {et~al.}(2007)\citenamefont {Bender},
  \citenamefont {Brody},\ and\ \citenamefont {Jones}}]{bender2007faster}%
  \BibitemOpen
  \bibfield  {author} {\bibinfo {author} {\bibfnamefont {C.~M.}\ \bibnamefont
  {Bender}}, \bibinfo {author} {\bibfnamefont {D.~C.}\ \bibnamefont {Brody}}, \
  and\ \bibinfo {author} {\bibfnamefont {H.~F.}\ \bibnamefont {Jones}},\ }\href
  {\doibase 10.1103/PhysRevLett.98.040403} {\bibfield  {journal} {\bibinfo
  {journal} {Physical Review Letters}\ }\textbf {\bibinfo {volume} {98}},\
  \bibinfo {pages} {040403} (\bibinfo {year} {2007})}\BibitemShut {NoStop}%
\bibitem [{\citenamefont {Lu}\ \emph {et~al.}()\citenamefont {Lu},
  \citenamefont {Liu}, \citenamefont {Liu}, \citenamefont {Rao}, \citenamefont
  {Lao}, \citenamefont {Wu}, \citenamefont {Zhu},\ and\ \citenamefont
  {Luo}}]{lu2022realizing}%
  \BibitemOpen
  \bibfield  {author} {\bibinfo {author} {\bibfnamefont {P.}~\bibnamefont
  {Lu}}, \bibinfo {author} {\bibfnamefont {T.}~\bibnamefont {Liu}}, \bibinfo
  {author} {\bibfnamefont {Y.}~\bibnamefont {Liu}}, \bibinfo {author}
  {\bibfnamefont {X.}~\bibnamefont {Rao}}, \bibinfo {author} {\bibfnamefont
  {Q.}~\bibnamefont {Lao}}, \bibinfo {author} {\bibfnamefont {H.}~\bibnamefont
  {Wu}}, \bibinfo {author} {\bibfnamefont {F.}~\bibnamefont {Zhu}}, \ and\
  \bibinfo {author} {\bibfnamefont {L.}~\bibnamefont {Luo}},\ }\href@noop {}
  {}\Eprint {http://arxiv.org/abs/2206.00940} {arXiv:2206.00940} \BibitemShut
  {NoStop}%
\bibitem [{\citenamefont {Li}\ \emph {et~al.}(2023)\citenamefont {Li},
  \citenamefont {Chen}, \citenamefont {Abbasi}, \citenamefont {Murch},\ and\
  \citenamefont {Whaley}}]{li2023speeding}%
  \BibitemOpen
  \bibfield  {author} {\bibinfo {author} {\bibfnamefont {Z.-Z.}\ \bibnamefont
  {Li}}, \bibinfo {author} {\bibfnamefont {W.}~\bibnamefont {Chen}}, \bibinfo
  {author} {\bibfnamefont {M.}~\bibnamefont {Abbasi}}, \bibinfo {author}
  {\bibfnamefont {K.~W.}\ \bibnamefont {Murch}}, \ and\ \bibinfo {author}
  {\bibfnamefont {K.~B.}\ \bibnamefont {Whaley}},\ }\href {\doibase
  10.1103/PhysRevLett.131.100202} {\bibfield  {journal} {\bibinfo  {journal}
  {Physical Review Letters}\ }\textbf {\bibinfo {volume} {131}},\ \bibinfo
  {pages} {100202} (\bibinfo {year} {2023})}\BibitemShut {NoStop}%
\bibitem [{\citenamefont {Chakraborty}\ and\ \citenamefont
  {Sarma}(2019)}]{chakraborty2019delayed}%
  \BibitemOpen
  \bibfield  {author} {\bibinfo {author} {\bibfnamefont {S.}~\bibnamefont
  {Chakraborty}}\ and\ \bibinfo {author} {\bibfnamefont {A.~K.}\ \bibnamefont
  {Sarma}},\ }\href {\doibase 10.1103/PhysRevA.100.063846} {\bibfield
  {journal} {\bibinfo  {journal} {Physical Review A}\ }\textbf {\bibinfo
  {volume} {100}},\ \bibinfo {pages} {063846} (\bibinfo {year}
  {2019})}\BibitemShut {NoStop}%
\bibitem [{\citenamefont {Fring}\ and\ \citenamefont
  {Frith}(2019)}]{fring2019eternal}%
  \BibitemOpen
  \bibfield  {author} {\bibinfo {author} {\bibfnamefont {A.}~\bibnamefont
  {Fring}}\ and\ \bibinfo {author} {\bibfnamefont {T.}~\bibnamefont {Frith}},\
  }\href {\doibase 10.1103/physreva.100.010102} {\bibfield  {journal} {\bibinfo
   {journal} {Physical Review A}\ }\textbf {\bibinfo {volume} {100}} (\bibinfo
  {year} {2019}),\ 10.1103/physreva.100.010102}\BibitemShut {NoStop}%
\bibitem [{\citenamefont {Hodaei}\ \emph {et~al.}(2017)\citenamefont {Hodaei},
  \citenamefont {Hassan}, \citenamefont {Wittek}, \citenamefont
  {Garcia-Gracia}, \citenamefont {El-Ganainy}, \citenamefont
  {Christodoulides},\ and\ \citenamefont {Khajavikhan}}]{hodaei2017enhanced}%
  \BibitemOpen
  \bibfield  {author} {\bibinfo {author} {\bibfnamefont {H.}~\bibnamefont
  {Hodaei}}, \bibinfo {author} {\bibfnamefont {A.~U.}\ \bibnamefont {Hassan}},
  \bibinfo {author} {\bibfnamefont {S.}~\bibnamefont {Wittek}}, \bibinfo
  {author} {\bibfnamefont {H.}~\bibnamefont {Garcia-Gracia}}, \bibinfo {author}
  {\bibfnamefont {R.}~\bibnamefont {El-Ganainy}}, \bibinfo {author}
  {\bibfnamefont {D.~N.}\ \bibnamefont {Christodoulides}}, \ and\ \bibinfo
  {author} {\bibfnamefont {M.}~\bibnamefont {Khajavikhan}},\ }\href {\doibase
  10.1038/nature23280} {\bibfield  {journal} {\bibinfo  {journal} {Nature}\
  }\textbf {\bibinfo {volume} {548}},\ \bibinfo {pages} {187} (\bibinfo {year}
  {2017})}\BibitemShut {NoStop}%
\bibitem [{\citenamefont {Chen}\ \emph {et~al.}(2017)\citenamefont {Chen},
  \citenamefont {{\"O}zdemir}, \citenamefont {Zhao}, \citenamefont {Wiersig},\
  and\ \citenamefont {Yang}}]{chen2017EP}%
  \BibitemOpen
  \bibfield  {author} {\bibinfo {author} {\bibfnamefont {W.}~\bibnamefont
  {Chen}}, \bibinfo {author} {\bibfnamefont {{\c S}.~K.}\ \bibnamefont
  {{\"O}zdemir}}, \bibinfo {author} {\bibfnamefont {G.}~\bibnamefont {Zhao}},
  \bibinfo {author} {\bibfnamefont {J.}~\bibnamefont {Wiersig}}, \ and\
  \bibinfo {author} {\bibfnamefont {L.}~\bibnamefont {Yang}},\ }\href {\doibase
  10.1038/nature23281} {\bibfield  {journal} {\bibinfo  {journal} {Nature}\
  }\textbf {\bibinfo {volume} {548}},\ \bibinfo {pages} {192} (\bibinfo {year}
  {2017})}\BibitemShut {NoStop}%
\bibitem [{\citenamefont {Parto}\ \emph {et~al.}(2025)\citenamefont {Parto},
  \citenamefont {Leefmans}, \citenamefont {Williams}, \citenamefont {Gray},\
  and\ \citenamefont {Marandi}}]{parto2025enhanced}%
  \BibitemOpen
  \bibfield  {author} {\bibinfo {author} {\bibfnamefont {M.}~\bibnamefont
  {Parto}}, \bibinfo {author} {\bibfnamefont {C.}~\bibnamefont {Leefmans}},
  \bibinfo {author} {\bibfnamefont {J.}~\bibnamefont {Williams}}, \bibinfo
  {author} {\bibfnamefont {R.~M.}\ \bibnamefont {Gray}}, \ and\ \bibinfo
  {author} {\bibfnamefont {A.}~\bibnamefont {Marandi}},\ }\href {\doibase
  10.1038/s41377-024-01667-z} {\bibfield  {journal} {\bibinfo  {journal}
  {Light: Science {\&} Applications}\ }\textbf {\bibinfo {volume} {14}},\
  \bibinfo {pages} {6} (\bibinfo {year} {2025})}\BibitemShut {NoStop}%
\bibitem [{\citenamefont {Wu}\ \emph {et~al.}(2025)\citenamefont {Wu},
  \citenamefont {Zhou}, \citenamefont {Liu}, \citenamefont {Kang},
  \citenamefont {Su},\ and\ \citenamefont {Yang}}]{wu2025enhanced}%
  \BibitemOpen
  \bibfield  {author} {\bibinfo {author} {\bibfnamefont {Q.-C.}\ \bibnamefont
  {Wu}}, \bibinfo {author} {\bibfnamefont {Y.-H.}\ \bibnamefont {Zhou}},
  \bibinfo {author} {\bibfnamefont {T.}~\bibnamefont {Liu}}, \bibinfo {author}
  {\bibfnamefont {Y.-H.}\ \bibnamefont {Kang}}, \bibinfo {author}
  {\bibfnamefont {Q.-P.}\ \bibnamefont {Su}}, \ and\ \bibinfo {author}
  {\bibfnamefont {C.-P.}\ \bibnamefont {Yang}},\ }\href {\doibase
  10.1016/j.cjph.2025.11.022} {\bibfield  {journal} {\bibinfo  {journal}
  {Chinese Journal of Physics}\ }\textbf {\bibinfo {volume} {98}},\ \bibinfo
  {pages} {1116–1129} (\bibinfo {year} {2025})}\BibitemShut {NoStop}%
\bibitem [{\citenamefont {Bernstein}\ and\ \citenamefont
  {Vazirani}(1997)}]{bernsteinQuantumComplexityTheory1997}%
  \BibitemOpen
  \bibfield  {author} {\bibinfo {author} {\bibfnamefont {E.}~\bibnamefont
  {Bernstein}}\ and\ \bibinfo {author} {\bibfnamefont {U.}~\bibnamefont
  {Vazirani}},\ }\href {https://doi.org/10.1137/S0097539796300921} {\bibfield
  {journal} {\bibinfo  {journal} {SIAM J. Comput.}\ }\textbf {\bibinfo {volume}
  {26}},\ \bibinfo {pages} {1411} (\bibinfo {year} {1997})}\BibitemShut
  {NoStop}%
\bibitem [{\citenamefont {Zhang}\ and\ \citenamefont
  {Wu}(2026)}]{zhang2026power}%
  \BibitemOpen
  \bibfield  {author} {\bibinfo {author} {\bibfnamefont {Q.}~\bibnamefont
  {Zhang}}\ and\ \bibinfo {author} {\bibfnamefont {B.}~\bibnamefont {Wu}},\
  }\href {\doibase 10.3390/e28030266} {\bibfield  {journal} {\bibinfo
  {journal} {Entropy}\ }\textbf {\bibinfo {volume} {28}} (\bibinfo {year}
  {2026}),\ 10.3390/e28030266}\BibitemShut {NoStop}%
\bibitem [{\citenamefont {Zhang}\ and\ \citenamefont
  {Wu}(2025)}]{zhang2025physics}%
  \BibitemOpen
  \bibfield  {author} {\bibinfo {author} {\bibfnamefont {Q.}~\bibnamefont
  {Zhang}}\ and\ \bibinfo {author} {\bibfnamefont {B.}~\bibnamefont {Wu}},\
  }\href {https://arxiv.org/abs/2506.18012} {\enquote {\bibinfo {title}
  {Physics and computation: A perspective from non-hermitian quantum
  computer},}\ } (\bibinfo {year} {2025}),\ \Eprint
  {http://arxiv.org/abs/2506.18012} {arXiv:2506.18012 [quant-ph]} \BibitemShut
  {NoStop}%
\bibitem [{\citenamefont {Toda}(1991)}]{toda1991}%
  \BibitemOpen
  \bibfield  {author} {\bibinfo {author} {\bibfnamefont {S.}~\bibnamefont
  {Toda}},\ }\href {\doibase 10.1137/0220053} {\bibfield  {journal} {\bibinfo
  {journal} {SIAM J. Comput.}\ }\textbf {\bibinfo {volume} {20}},\ \bibinfo
  {pages} {865} (\bibinfo {year} {1991})}\BibitemShut {NoStop}%
\bibitem [{\citenamefont {Brun}(2002)}]{brunsimplemodelquantum2002}%
  \BibitemOpen
  \bibfield  {author} {\bibinfo {author} {\bibfnamefont {T.~A.}\ \bibnamefont
  {Brun}},\ }\href {\doibase 10.1119/1.1475328} {\bibfield  {journal} {\bibinfo
   {journal} {American Journal of Physics}\ }\textbf {\bibinfo {volume} {70}},\
  \bibinfo {pages} {719} (\bibinfo {year} {2002})}\BibitemShut {NoStop}%
\bibitem [{\citenamefont
  {Lindblad}(1976)}]{lindbladGeneratorsQuantumDynamical1976}%
  \BibitemOpen
  \bibfield  {author} {\bibinfo {author} {\bibfnamefont {G.}~\bibnamefont
  {Lindblad}},\ }\href {\doibase 10.1007/BF01608499} {\bibfield  {journal}
  {\bibinfo  {journal} {Communications in Mathematical Physics}\ }\textbf
  {\bibinfo {volume} {48}},\ \bibinfo {pages} {119} (\bibinfo {year}
  {1976})}\BibitemShut {NoStop}%
\bibitem [{\citenamefont {Alicki}\ and\ \citenamefont
  {Lendi}(2007)}]{alickiQuantumDynamicalSemigroups2007}%
  \BibitemOpen
  \bibfield  {author} {\bibinfo {author} {\bibfnamefont {R.}~\bibnamefont
  {Alicki}}\ and\ \bibinfo {author} {\bibfnamefont {K.}~\bibnamefont {Lendi}},\
  }\href@noop {} {\emph {\bibinfo {title} {Quantum Dynamical Semigroups and
  Applications}}},\ \bibinfo {series} {Lecture Notes in Physics}\ No.\ \bibinfo
  {number} {717}\ (\bibinfo  {publisher} {{Springer-Verlag}},\ \bibinfo
  {address} {{Berlin ; New York}},\ \bibinfo {year} {2007})\BibitemShut
  {NoStop}%
\bibitem [{\citenamefont {Breuer}\ and\ \citenamefont
  {Petruccione}(2002)}]{breuerTheoryOpenQuantum2002}%
  \BibitemOpen
  \bibfield  {author} {\bibinfo {author} {\bibfnamefont {H.-P.}\ \bibnamefont
  {Breuer}}\ and\ \bibinfo {author} {\bibfnamefont {F.}~\bibnamefont
  {Petruccione}},\ }\href@noop {} {\emph {\bibinfo {title} {The Theory of Open
  Quantum Systems}}}\ (\bibinfo  {publisher} {{Oxford University Press}},\
  \bibinfo {address} {{Oxford ; New York}},\ \bibinfo {year}
  {2002})\BibitemShut {NoStop}%
\bibitem [{\citenamefont {Karuvade}\ \emph {et~al.}(2022)\citenamefont
  {Karuvade}, \citenamefont {Alase},\ and\ \citenamefont
  {Sanders}}]{karuvade2022observing}%
  \BibitemOpen
  \bibfield  {author} {\bibinfo {author} {\bibfnamefont {S.}~\bibnamefont
  {Karuvade}}, \bibinfo {author} {\bibfnamefont {A.}~\bibnamefont {Alase}}, \
  and\ \bibinfo {author} {\bibfnamefont {B.~C.}\ \bibnamefont {Sanders}},\
  }\href {\doibase 10.1103/PhysRevResearch.4.013016} {\bibfield  {journal}
  {\bibinfo  {journal} {Phys. Rev. Res.}\ }\textbf {\bibinfo {volume} {4}},\
  \bibinfo {pages} {013016} (\bibinfo {year} {2022})}\BibitemShut {NoStop}%
\bibitem [{\citenamefont {Papadimitriou}(1995)}]{Papadimitriou:book}%
  \BibitemOpen
  \bibfield  {author} {\bibinfo {author} {\bibfnamefont {C.}~\bibnamefont
  {Papadimitriou}},\ }\href@noop {} {\emph {\bibinfo {title} {Computational
  Complexity}}}\ (\bibinfo  {publisher} {Addison Wesley Longman},\ \bibinfo
  {address} {Reading, Massachusetts},\ \bibinfo {year} {1995})\BibitemShut
  {NoStop}%
\bibitem [{\citenamefont {{A.Yu. Kitaev, A.H. Shen, M.N.
  Vyalyi}}(2000)}]{Kitaev:book}%
  \BibitemOpen
  \bibfield  {author} {\bibinfo {author} {\bibnamefont {{A.Yu. Kitaev, A.H.
  Shen, M.N. Vyalyi}}},\ }\href@noop {} {\emph {\bibinfo {title} {Classical and
  Quantum Computation}}},\ \bibinfo {series} {Graduate Studies in Mathematics},
  Vol.~\bibinfo {volume} {47}\ (\bibinfo  {publisher} {American Mathematical
  Society},\ \bibinfo {address} {Providence, RI},\ \bibinfo {year}
  {2000})\BibitemShut {NoStop}%
\bibitem [{\citenamefont {Arora}\ and\ \citenamefont
  {Barak}(2009)}]{arora_computational_2009}%
  \BibitemOpen
  \bibfield  {author} {\bibinfo {author} {\bibfnamefont {S.}~\bibnamefont
  {Arora}}\ and\ \bibinfo {author} {\bibfnamefont {B.}~\bibnamefont {Barak}},\
  }\href {\doibase 10.1017/CBO9780511804090} {\emph {\bibinfo {title}
  {Computational {{Complexity}}: {{A Modern Approach}}}}}\ (\bibinfo
  {publisher} {{Cambridge University Press}},\ \bibinfo {address}
  {{Cambridge}},\ \bibinfo {year} {2009})\BibitemShut {NoStop}%
\bibitem [{\citenamefont {Aaronson}\ and\ \citenamefont
  {Arkhipov}(2011)}]{boson-sampling-orig}%
  \BibitemOpen
  \bibfield  {author} {\bibinfo {author} {\bibfnamefont {S.}~\bibnamefont
  {Aaronson}}\ and\ \bibinfo {author} {\bibfnamefont {A.}~\bibnamefont
  {Arkhipov}},\ }in\ \href {\doibase 10.1145/1993636.1993682} {\emph {\bibinfo
  {booktitle} {Proceedings of the Forty-Third Annual ACM Symposium on Theory of
  Computing}}},\ \bibinfo {series and number} {STOC '11}\ (\bibinfo
  {publisher} {Association for Computing Machinery},\ \bibinfo {address} {New
  York, NY, USA},\ \bibinfo {year} {2011})\ pp.\ \bibinfo {pages}
  {333--342}\BibitemShut {NoStop}%
\bibitem [{\citenamefont {Aaronson}(2014)}]{aaronson2014postbqp}%
  \BibitemOpen
  \bibfield  {author} {\bibinfo {author} {\bibfnamefont {S.}~\bibnamefont
  {Aaronson}},\ }\href@noop {} {\enquote {\bibinfo {title} {{PostBQP
  Postscripts: A Confession of Mathematical Errors}},}\ }\bibinfo
  {howpublished} {\url{https://scottaaronson.blog/?p=2072}} (\bibinfo {year}
  {2014}),\ \bibinfo {note} {blog post on Shtetl-Optimized}\BibitemShut
  {NoStop}%
\bibitem [{\citenamefont {Aaronson}\ and\ \citenamefont
  {Ambainis}(2015)}]{aaronson2014forrelation}%
  \BibitemOpen
  \bibfield  {author} {\bibinfo {author} {\bibfnamefont {S.}~\bibnamefont
  {Aaronson}}\ and\ \bibinfo {author} {\bibfnamefont {A.}~\bibnamefont
  {Ambainis}},\ }in\ \href {\doibase 10.1145/2746539.2746547} {\emph {\bibinfo
  {booktitle} {Proceedings of the Forty-Seventh Annual ACM Symposium on Theory
  of Computing}}},\ \bibinfo {series and number} {STOC '15}\ (\bibinfo
  {publisher} {Association for Computing Machinery},\ \bibinfo {address} {New
  York, NY, USA},\ \bibinfo {year} {2015})\ pp.\ \bibinfo {pages}
  {307--316}\BibitemShut {NoStop}%
\bibitem [{\citenamefont {Aaronson}\ and\ \citenamefont
  {Chen}(2017)}]{aaronson2016}%
  \BibitemOpen
  \bibfield  {author} {\bibinfo {author} {\bibfnamefont {S.}~\bibnamefont
  {Aaronson}}\ and\ \bibinfo {author} {\bibfnamefont {L.}~\bibnamefont
  {Chen}},\ }in\ \href {https://arxiv.org/abs/1612.05903} {\emph {\bibinfo
  {booktitle} {Proceedings of the 32nd Computational Complexity Conference}}},\
  \bibinfo {series and number} {CCC '17}\ (\bibinfo  {publisher} {Schloss
  Dagstuhl--Leibniz-Zentrum fuer Informatik},\ \bibinfo {address} {Dagstuhl,
  DEU},\ \bibinfo {year} {2017})\BibitemShut {NoStop}%
\bibitem [{\citenamefont {Chen}(2016)}]{chen2016note}%
  \BibitemOpen
  \bibfield  {author} {\bibinfo {author} {\bibfnamefont {L.}~\bibnamefont
  {Chen}},\ }\href {https://arxiv.org/abs/1605.00619} {\enquote {\bibinfo
  {title} {A note on oracle separations for bqp},}\ } (\bibinfo {year}
  {2016}),\ \Eprint {http://arxiv.org/abs/1605.00619} {arXiv:1605.00619
  [quant-ph]} \BibitemShut {NoStop}%
\bibitem [{\citenamefont {Bremner}\ \emph {et~al.}(2011)\citenamefont
  {Bremner}, \citenamefont {Jozsa},\ and\ \citenamefont
  {Shepherd}}]{bremner2011classical}%
  \BibitemOpen
  \bibfield  {author} {\bibinfo {author} {\bibfnamefont {M.~J.}\ \bibnamefont
  {Bremner}}, \bibinfo {author} {\bibfnamefont {R.}~\bibnamefont {Jozsa}}, \
  and\ \bibinfo {author} {\bibfnamefont {D.~J.}\ \bibnamefont {Shepherd}},\
  }\href@noop {} {\bibfield  {journal} {\bibinfo  {journal} {Proceedings of the
  Royal Society A: Mathematical, Physical and Engineering Sciences}\ }\textbf
  {\bibinfo {volume} {467}},\ \bibinfo {pages} {459} (\bibinfo {year}
  {2011})}\BibitemShut {NoStop}%
\bibitem [{\citenamefont {Van Den~Nes}(2010)}]{nes2010classical}%
  \BibitemOpen
  \bibfield  {author} {\bibinfo {author} {\bibfnamefont {M.}~\bibnamefont {Van
  Den~Nes}},\ }\href@noop {} {\bibfield  {journal} {\bibinfo  {journal}
  {Quantum Info. Comput.}\ }\textbf {\bibinfo {volume} {10}},\ \bibinfo {pages}
  {258–271} (\bibinfo {year} {2010})}\BibitemShut {NoStop}%
\bibitem [{\citenamefont {Johnson}\ \emph {et~al.}(2013)\citenamefont
  {Johnson}, \citenamefont {Biamonte}, \citenamefont {Clark},\ and\
  \citenamefont {Jaksch}}]{johnson2013solving}%
  \BibitemOpen
  \bibfield  {author} {\bibinfo {author} {\bibfnamefont {T.~H.}\ \bibnamefont
  {Johnson}}, \bibinfo {author} {\bibfnamefont {J.~D.}\ \bibnamefont
  {Biamonte}}, \bibinfo {author} {\bibfnamefont {S.~R.}\ \bibnamefont {Clark}},
  \ and\ \bibinfo {author} {\bibfnamefont {D.}~\bibnamefont {Jaksch}},\ }\href
  {\doibase 10.1038/srep01235} {\bibfield  {journal} {\bibinfo  {journal}
  {Scientific Reports}\ }\textbf {\bibinfo {volume} {3}} (\bibinfo {year}
  {2013}),\ 10.1038/srep01235}\BibitemShut {NoStop}%
\bibitem [{\citenamefont {Osborne}(2006)}]{osborne2006efficient}%
  \BibitemOpen
  \bibfield  {author} {\bibinfo {author} {\bibfnamefont {T.~J.}\ \bibnamefont
  {Osborne}},\ }\href {\doibase 10.1103/PhysRevLett.97.157202} {\bibfield
  {journal} {\bibinfo  {journal} {Phys. Rev. Lett.}\ }\textbf {\bibinfo
  {volume} {97}},\ \bibinfo {pages} {157202} (\bibinfo {year}
  {2006})}\BibitemShut {NoStop}%
\bibitem [{\citenamefont {Hastings}(2009)}]{hastings2009quantum}%
  \BibitemOpen
  \bibfield  {author} {\bibinfo {author} {\bibfnamefont {M.~B.}\ \bibnamefont
  {Hastings}},\ }\href {\doibase 10.1103/physrevlett.103.050502} {\bibfield
  {journal} {\bibinfo  {journal} {Physical Review Letters}\ }\textbf {\bibinfo
  {volume} {103}} (\bibinfo {year} {2009}),\
  10.1103/physrevlett.103.050502}\BibitemShut {NoStop}%
\bibitem [{\citenamefont {Aloisio}\ \emph {et~al.}(2023)\citenamefont
  {Aloisio}, \citenamefont {White}, \citenamefont {Hill},\ and\ \citenamefont
  {Modi}}]{aloisio2023sampling}%
  \BibitemOpen
  \bibfield  {author} {\bibinfo {author} {\bibfnamefont {I.}~\bibnamefont
  {Aloisio}}, \bibinfo {author} {\bibfnamefont {G.}~\bibnamefont {White}},
  \bibinfo {author} {\bibfnamefont {C.}~\bibnamefont {Hill}}, \ and\ \bibinfo
  {author} {\bibfnamefont {K.}~\bibnamefont {Modi}},\ }\href {\doibase
  10.1103/PRXQuantum.4.020310} {\bibfield  {journal} {\bibinfo  {journal} {PRX
  Quantum}\ }\textbf {\bibinfo {volume} {4}},\ \bibinfo {pages} {020310}
  (\bibinfo {year} {2023})}\BibitemShut {NoStop}%
\bibitem [{\citenamefont {Aaronson}(2011)}]{aaronson2010equivalence}%
  \BibitemOpen
  \bibfield  {author} {\bibinfo {author} {\bibfnamefont {S.}~\bibnamefont
  {Aaronson}},\ }in\ \href
  {https://link.springer.com/chapter/10.1007/978-3-642-20712-9_1#citeas} {\emph
  {\bibinfo {booktitle} {Computer Science -- Theory and Applications}}},\
  \bibinfo {editor} {edited by\ \bibinfo {editor} {\bibfnamefont
  {A.}~\bibnamefont {Kulikov}}\ and\ \bibinfo {editor} {\bibfnamefont
  {N.}~\bibnamefont {Vereshchagin}}}\ (\bibinfo  {publisher} {Springer Berlin
  Heidelberg},\ \bibinfo {address} {Berlin, Heidelberg},\ \bibinfo {year}
  {2011})\ pp.\ \bibinfo {pages} {1--14}\BibitemShut {NoStop}%
\bibitem [{\citenamefont {Kuperberg}(2009)}]{kuperberg_how_2014}%
  \BibitemOpen
  \bibfield  {author} {\bibinfo {author} {\bibfnamefont {G.}~\bibnamefont
  {Kuperberg}},\ }\href {\doibase 10.4086/toc.2015.v011a006} {\bibfield
  {journal} {\bibinfo  {journal} {Theory of Computing}\ }\textbf {\bibinfo
  {volume} {11}} (\bibinfo {year} {2009}),\
  10.4086/toc.2015.v011a006}\BibitemShut {NoStop}%
\bibitem [{\citenamefont {Adleman}\ \emph {et~al.}(1997)\citenamefont
  {Adleman}, \citenamefont {DeMarrais},\ and\ \citenamefont
  {Huang}}]{adleman1997quantum}%
  \BibitemOpen
  \bibfield  {author} {\bibinfo {author} {\bibfnamefont {L.~M.}\ \bibnamefont
  {Adleman}}, \bibinfo {author} {\bibfnamefont {J.}~\bibnamefont {DeMarrais}},
  \ and\ \bibinfo {author} {\bibfnamefont {M.-D.~A.}\ \bibnamefont {Huang}},\
  }\href {\doibase 10.1137/S0097539795293639} {\bibfield  {journal} {\bibinfo
  {journal} {SIAM Journal on Computing}\ }\textbf {\bibinfo {volume} {26}},\
  \bibinfo {pages} {1524} (\bibinfo {year} {1997})},\ \Eprint
  {http://arxiv.org/abs/https://doi.org/10.1137/S0097539795293639}
  {https://doi.org/10.1137/S0097539795293639} \BibitemShut {NoStop}%
\bibitem [{\citenamefont {Nielsen}\ and\ \citenamefont
  {Chuang}(2010)}]{nielsen_quantum_2010}%
  \BibitemOpen
  \bibfield  {author} {\bibinfo {author} {\bibfnamefont {M.~A.}\ \bibnamefont
  {Nielsen}}\ and\ \bibinfo {author} {\bibfnamefont {I.~L.}\ \bibnamefont
  {Chuang}},\ }\href@noop {} {\emph {\bibinfo {title} {Quantum Computation and
  Quantum Information}}},\ \bibinfo {edition} {10th}\ ed.\ (\bibinfo
  {publisher} {{Cambridge University Press}},\ \bibinfo {address} {{Cambridge ;
  New York}},\ \bibinfo {year} {2010})\BibitemShut {NoStop}%
\bibitem [{\citenamefont {Gottesman}(1998)}]{gottesman1998heisenberg}%
  \BibitemOpen
  \bibfield  {author} {\bibinfo {author} {\bibfnamefont {D.}~\bibnamefont
  {Gottesman}},\ }\href {https://arxiv.org/abs/quant-ph/9807006} {\enquote
  {\bibinfo {title} {The heisenberg representation of quantum computers},}\ }
  (\bibinfo {year} {1998}),\ \Eprint {http://arxiv.org/abs/quant-ph/9807006}
  {arXiv:quant-ph/9807006 [quant-ph]} \BibitemShut {NoStop}%
\bibitem [{\citenamefont {Jozsa}\ and\ \citenamefont
  {Nest}(2013)}]{jozsa2013classical}%
  \BibitemOpen
  \bibfield  {author} {\bibinfo {author} {\bibfnamefont {R.}~\bibnamefont
  {Jozsa}}\ and\ \bibinfo {author} {\bibfnamefont {M.}~\bibnamefont {Nest}},\
  }\href {\doibase 10.26421/QIC14.7-8-7} {\bibfield  {journal} {\bibinfo
  {journal} {Quantum Information and Computation}\ }\textbf {\bibinfo {volume}
  {14}} (\bibinfo {year} {2013}),\ 10.26421/QIC14.7-8-7}\BibitemShut {NoStop}%
\bibitem [{\citenamefont {Aaronson}\ and\ \citenamefont
  {Gottesman}(2004)}]{Aaronson:2004aa}%
  \BibitemOpen
  \bibfield  {author} {\bibinfo {author} {\bibfnamefont {S.}~\bibnamefont
  {Aaronson}}\ and\ \bibinfo {author} {\bibfnamefont {D.}~\bibnamefont
  {Gottesman}},\ }\href {\doibase 10.1103/PhysRevA.70.052328} {\bibfield
  {journal} {\bibinfo  {journal} {Physical Review A}\ }\textbf {\bibinfo
  {volume} {70}},\ \bibinfo {pages} {052328} (\bibinfo {year}
  {2004})}\BibitemShut {NoStop}%
\bibitem [{\citenamefont {Sang}\ \emph {et~al.}(2023)\citenamefont {Sang},
  \citenamefont {Li}, \citenamefont {Hsieh},\ and\ \citenamefont
  {Yoshida}}]{sang2023ultrafrast}%
  \BibitemOpen
  \bibfield  {author} {\bibinfo {author} {\bibfnamefont {S.}~\bibnamefont
  {Sang}}, \bibinfo {author} {\bibfnamefont {Z.}~\bibnamefont {Li}}, \bibinfo
  {author} {\bibfnamefont {T.~H.}\ \bibnamefont {Hsieh}}, \ and\ \bibinfo
  {author} {\bibfnamefont {B.}~\bibnamefont {Yoshida}},\ }\href
  {https://link.aps.org/doi/10.1103/PRXQuantum.4.040332} {\bibfield  {journal}
  {\bibinfo  {journal} {PRX Quantum}\ }\textbf {\bibinfo {volume} {4}},\
  \bibinfo {pages} {040332} (\bibinfo {year} {2023})}\BibitemShut {NoStop}%
\bibitem [{\citenamefont {{L.G. Valiant}}(2002)}]{valiant}%
  \BibitemOpen
  \bibfield  {author} {\bibinfo {author} {\bibnamefont {{L.G. Valiant}}},\
  }\href@noop {} {\bibfield  {journal} {\bibinfo  {journal} {{SIAM J. on
  Computing}}\ }\textbf {\bibinfo {volume} {31}},\ \bibinfo {pages} {1229}
  (\bibinfo {year} {2002})}\BibitemShut {NoStop}%
\bibitem [{\citenamefont {Jozsa}\ and\ \citenamefont
  {Miyake}(2008)}]{jozsa2008matchgates}%
  \BibitemOpen
  \bibfield  {author} {\bibinfo {author} {\bibfnamefont {R.}~\bibnamefont
  {Jozsa}}\ and\ \bibinfo {author} {\bibfnamefont {A.}~\bibnamefont {Miyake}},\
  }\href {\doibase 10.1098/rspa.2008.0189} {\bibfield  {journal} {\bibinfo
  {journal} {Proceedings of the Royal Society A: Mathematical, Physical and
  Engineering Sciences}\ }\textbf {\bibinfo {volume} {464}},\ \bibinfo {pages}
  {3089} (\bibinfo {year} {2008})}\BibitemShut {NoStop}%
\bibitem [{\citenamefont {Shtanko}\ \emph {et~al.}(2021)\citenamefont
  {Shtanko}, \citenamefont {Deshpande}, \citenamefont {Julienne},\ and\
  \citenamefont {Gorshkov}}]{shtanko2021complexity}%
  \BibitemOpen
  \bibfield  {author} {\bibinfo {author} {\bibfnamefont {O.}~\bibnamefont
  {Shtanko}}, \bibinfo {author} {\bibfnamefont {A.}~\bibnamefont {Deshpande}},
  \bibinfo {author} {\bibfnamefont {P.~S.}\ \bibnamefont {Julienne}}, \ and\
  \bibinfo {author} {\bibfnamefont {A.~V.}\ \bibnamefont {Gorshkov}},\ }\href
  {\doibase 10.1103/PRXQuantum.2.030350} {\bibfield  {journal} {\bibinfo
  {journal} {PRX Quantum}\ }\textbf {\bibinfo {volume} {2}},\ \bibinfo {pages}
  {030350} (\bibinfo {year} {2021})}\BibitemShut {NoStop}%
\bibitem [{\citenamefont {Projansky}\ \emph {et~al.}(2025)\citenamefont
  {Projansky}, \citenamefont {Necaise},\ and\ \citenamefont
  {Whitfield}}]{projansky2025gaussianity}%
  \BibitemOpen
  \bibfield  {author} {\bibinfo {author} {\bibfnamefont {A.~M.}\ \bibnamefont
  {Projansky}}, \bibinfo {author} {\bibfnamefont {J.}~\bibnamefont {Necaise}},
  \ and\ \bibinfo {author} {\bibfnamefont {J.~D.}\ \bibnamefont {Whitfield}},\
  }\href {\doibase 10.1088/1751-8121/adcd15} {\bibfield  {journal} {\bibinfo
  {journal} {Journal of Physics A: Mathematical and Theoretical}\ }\textbf
  {\bibinfo {volume} {58}},\ \bibinfo {pages} {195302} (\bibinfo {year}
  {2025})}\BibitemShut {NoStop}%
\bibitem [{\citenamefont {Cleve}\ and\ \citenamefont
  {Wang}(2017)}]{cleve2017efficient}%
  \BibitemOpen
  \bibfield  {author} {\bibinfo {author} {\bibfnamefont {R.}~\bibnamefont
  {Cleve}}\ and\ \bibinfo {author} {\bibfnamefont {C.}~\bibnamefont {Wang}},\
  }in\ \href {\doibase 10.4230/LIPIcs.ICALP.2017.17} {\emph {\bibinfo
  {booktitle} {44th International Colloquium on Automata, Languages, and
  Programming (ICALP 2017)}}},\ \bibinfo {series} {Leibniz International
  Proceedings in Informatics (LIPIcs)}, Vol.~\bibinfo {volume} {80},\ \bibinfo
  {editor} {edited by\ \bibinfo {editor} {\bibfnamefont {I.}~\bibnamefont
  {Chatzigiannakis}}, \bibinfo {editor} {\bibfnamefont {P.}~\bibnamefont
  {Indyk}}, \bibinfo {editor} {\bibfnamefont {F.}~\bibnamefont {Kuhn}}, \ and\
  \bibinfo {editor} {\bibfnamefont {A.}~\bibnamefont {Muscholl}}}\ (\bibinfo
  {publisher} {Schloss Dagstuhl -- Leibniz-Zentrum f{\"u}r Informatik},\
  \bibinfo {address} {Dagstuhl, Germany},\ \bibinfo {year} {2017})\ pp.\
  \bibinfo {pages} {17:1--17:14}\BibitemShut {NoStop}%
\bibitem [{\citenamefont {Ding}\ \emph {et~al.}(2024)\citenamefont {Ding},
  \citenamefont {Li},\ and\ \citenamefont {Lin}}]{ding2024simulating}%
  \BibitemOpen
  \bibfield  {author} {\bibinfo {author} {\bibfnamefont {Z.}~\bibnamefont
  {Ding}}, \bibinfo {author} {\bibfnamefont {X.}~\bibnamefont {Li}}, \ and\
  \bibinfo {author} {\bibfnamefont {L.}~\bibnamefont {Lin}},\ }\href {\doibase
  10.1103/PRXQuantum.5.020332} {\bibfield  {journal} {\bibinfo  {journal} {PRX
  Quantum}\ }\textbf {\bibinfo {volume} {5}},\ \bibinfo {pages} {020332}
  (\bibinfo {year} {2024})}\BibitemShut {NoStop}%
\bibitem [{\citenamefont {Paiva}\ \emph {et~al.}(2022)\citenamefont {Paiva},
  \citenamefont {Te'eni}, \citenamefont {Peled}, \citenamefont {Cohen},\ and\
  \citenamefont {Aharonov}}]{Paiva_2022}%
  \BibitemOpen
  \bibfield  {author} {\bibinfo {author} {\bibfnamefont {I.~L.}\ \bibnamefont
  {Paiva}}, \bibinfo {author} {\bibfnamefont {A.}~\bibnamefont {Te'eni}},
  \bibinfo {author} {\bibfnamefont {B.~Y.}\ \bibnamefont {Peled}}, \bibinfo
  {author} {\bibfnamefont {E.}~\bibnamefont {Cohen}}, \ and\ \bibinfo {author}
  {\bibfnamefont {Y.}~\bibnamefont {Aharonov}},\ }\href {\doibase
  10.1038/s42005-022-01081-0} {\bibfield  {journal} {\bibinfo  {journal}
  {Communications Physics}\ }\textbf {\bibinfo {volume} {5}} (\bibinfo {year}
  {2022}),\ 10.1038/s42005-022-01081-0}\BibitemShut {NoStop}%
\bibitem [{\citenamefont {Fan}\ and\ \citenamefont
  {Hoffman}(1955)}]{fan1955some}%
  \BibitemOpen
  \bibfield  {author} {\bibinfo {author} {\bibfnamefont {K.}~\bibnamefont
  {Fan}}\ and\ \bibinfo {author} {\bibfnamefont {A.~J.}\ \bibnamefont
  {Hoffman}},\ }\href {\doibase 10.1090/S0002-9939-1955-0067841-7} {\bibfield
  {journal} {\bibinfo  {journal} {Proceedings of the American Mathematical
  Society}\ }\textbf {\bibinfo {volume} {6}},\ \bibinfo {pages} {111} (\bibinfo
  {year} {1955})}\BibitemShut {NoStop}%
\end{thebibliography}%

\end{document}